\documentclass[12pt]{article}

\usepackage{hyperref}
\usepackage{epsf}
\usepackage{amsmath}
\usepackage{graphics}
\usepackage{amsfonts}
\usepackage{amssymb}
\usepackage{latexsym}
\usepackage{color}
\usepackage[dvips]{graphicx}
\input{colordvi.tex}

\begin{document}
\begin{titlepage}
\begin{flushright}
LMU-ASC 79/12,~RESCEU-50/12
\end{flushright}
\begin{center}

\vskip .5in

{\Large \bf
Theoretical Models of Dark Energy
}

\vskip .45in

{\large Jaewon Yoo }\footnote{ \texttt{jaewon.yoo `at' physik.uni-muenchen.de}}

\vskip .1in
{\em
  Arnold Sommerfeld Center for Theoretical Physics,\\
  Ludwig Maximilian University of Munich, Theresienstrasse 37, 80333 Munich, Germany
  }  

\vskip .4in
{\large Yuki Watanabe }\footnote{ \texttt{watanabe `at' resceu.s.u-tokyo.ac.jp}}
\vskip.1in
{\em
  Research Center for the Early Universe,\\ 
  University of Tokyo, Bunkyo-ku, Tokyo 113-0033, Japan
  }
  
\vskip.2in
(Dated: \today)
\end{center}

\vskip .4in

\begin{abstract}
Mounting observational data confirm that about 73\% of the energy density consists of dark energy which is responsible for the current accelerated expansion of the Universe. We present observational evidences and dark energy projects.
We then review various theoretical ideas that have been proposed to explain the origin of dark energy; they contain the cosmological constant, modified matter models, modified gravity models and the inhomogeneous model.
The cosmological constant suffers from two major problems: one regarding fine-tuning and the other regarding coincidence. To solve them there arose modified matter models such as quintessence, k-essence, coupled dark energy, and unified dark energy. We compare those models by presenting attractive aspects, new rising problems and possible solutions.
Furthermore we review modified gravity models that lead to late-time accelerated expansion without invoking a new form of dark energy; they contain f(R) gravity and the Dvali-Gabadadze-Porrati model. We also discuss observational constraints on those models and on future modified gravity theories.
Finally we review the inhomogeneous Lema\^{i}tre-Tolman-Bondi model that drops an assumption of the spatial homogeneity of the Universe.
We also present basics of cosmology and scalar field theory, which are useful especially for students and novices to understand dark energy models.
\end{abstract}
\end{titlepage}
\renewcommand{\thepage}{\arabic{page}}
\setcounter{page}{1}
\tableofcontents
\section{Introduction}\label{sec:intro}

 All physical bodies attract each other. Then why do our stars, galaxies and the Universe not collapse?
 Einstein believed the Universe is static, but his {\it Einstein equations} implied a dynamic Universe. In 1917 he introduced the cosmological constant $\Lambda$, the antigravity vacuum energy. 
 He had adjusted the scale of $\Lambda$ to gravity so that the Universe stays static. 
 In the same year 1917 Willem de Sitter has calculated Einstein's equation for a vacuum Universe without matter. This solution drives inflation of the Universe. Ironically, in Einstein's Universe $\Lambda$ was used for a stationary Universe, but in de Sitter's Universe $\Lambda$ is the reason for inflation.
 In 1928 Edwin Hubble demonstrated that the Universe is expanding and Einstein called the cosmological constant {\it the biggest blunder} of his lifetime. 
 
The discovery of Riess et al. in 1998 \cite{Riess:1998cb} and Perlmutter et al. in 1999 \cite{Perlmutter:1998np} changed the landscape of a vision of the Universe.
They found that distant supernovae at $z\sim0.5$ are $\Delta m\sim0.25$ mag, about 25\%, fainter than ones expected for a decelerating Universe without the cosmological constant. They concluded from the observation of supernovae that the Universe is currently accelerating in its expansion. This discovery won them the Nobel Prize in physics 2011.

We call the source of this acceleration {\it dark energy}. 
Not only Type Ia supernova (SN Ia) observations but also Cosmic Microwave Background (CMB) and Baryon Acoustic Oscillations (BAO) substantiate that dark energy captures about $73\%$ of the energy density of the current Universe.
Dark energy is a major outstanding issue in physics and cosmology today. There are a number of useful reviews of dark energy that mainly focused on theory \cite{Copeland:2006wr,Tsujikawa:2010sc,Tsujikawa:2010zza}, on probes of dark energy \cite{Frieman:2008sn} and on the cosmological constant \cite{Carroll:2000fy,Martin:2012bt}.

The aim of this review is to introduce major theoretical models of dark energy focusing on the basic physics of each model for students and non-experts on the subject. Since we do not cover all of models in the field, the list of references are limited. The interested readers should consult reviews with more complete reference lists, e.g. \cite{Tsujikawa:2010zza}. In the first section we present observational evidences (section~\ref{sec:intro}). 
In the following sections we introduce theoretical approaches to explain the origin of dark energy. In section~\ref{sec:cc}, we review the simplest candidate: the cosmological constant $\Lambda$; the cosmological constant suffers from two major problems, one regarding fine-tuning and the other one coincidence. To solve them there arose modified matter models such as quintessence (section~\ref{sec:quintessence}), k-essence (section~\ref{sec:kessence}), coupled dark energy (section~\ref{sec:coupledDE}), and unified models of dark energy and dark matter (section~\ref{sec:unifiedDE}). We compare those models by presenting attractive aspects, new rising problems and possible solutions. Furthermore, we review modified gravity models which lead to late-time accelerated expansion without invoking a new form of dark energy. They include $f(R)$ gravity (section~\ref{sec:f(R)}) and the DGP model (section~\ref{sec:DGP}). We also discuss observational constraints on these models. In section~\ref{sec:LTB} we review the inhomogeneous Lema\^{i}tre-Tolman-Bondi model that drops an assumption of the spatial homogeneity of the Universe. Lastly we summarize and give a brief comparison of those candidates. In the appendix we present basics of cosmology including the homogeneous FLRW model and scalar field theory, which are useful to understand dark energy models. 
The sections are connected with each other but independent enough for readers to look directly into their interested sections without referring to the preceding sections. 

We use the metric signature ($+,-,-,-$) and follow mostly the conventions of Mukhanov's book \cite{Mukhanov:2005sc}. We employ natural units $c=\hbar=k_B=1$ while retain the gravitational constant $G \equiv 1/M_{pl}^2$, where $M_{pl} = 2.177\times 10^{-5} {\rm g} = 1.416\times 10^{32} {\rm K}=1.221 \times 10^{19} {\rm GeV}$ is the Planck mass.

\subsection{Observational evidences on dark energy}

The recently dominating dark energy is supported by many independent observations, such as SN Ia \cite{Riess:2004nr}, CMB \cite{Bernardis:2000gy} and BAO \cite{Eisenstein:2005su}. Here we present how these could be evidences for the presence of dark energy.

\subsubsection*{Type Ia supernova (SN Ia)}
  
As we mentioned above, SN Ia observations are one significant evidence for the cosmic acceleration. When a white dwarf reaches $1.4 M_{\odot}$, gravity becomes bigger than the Fermi degeneracy pressure and drives the star into explosion. Since SN Ia is made in the same process, we assume that SN Ia always have the same luminosity and it can be used as a {\it standard candle}. Measuring the apparent magnitude, we can estimate the luminosity distance. A supernova is as bright as a whole galaxy, thus surveying SN Ia is very powerful and a direct way to study the distant Universe.

But since the detection was first published some scientists had doubt that the distant supernovae could appear fainter just due to extinction. Absorption by intervening dust could also lead to characteristic reddening. Nowadays  thanks to Hubble Space Telescope (HST) \cite{Knop:2003iy} we have high quality light curves up to $z \simeq 1.8$ and big amount of other distant SN data from ground based observations. 
We are thus more convinced that this was not due to extinction by dust but due to real cosmic acceleration.

\subsubsection*{Cosmic Microwave Background (CMB)}

Not only SN Ia but also CMB observations strongly indicate the presence of dark energy. The CMB is a snap shot of the Universe before the cosmic structure developed and has temperature anisotropies which are influenced by dark energy. This snap shot tells us the cosmic history from the photon decoupling epoch to the present.

The angular power spectrum of CMB temperature anisotropies measured by Wilkinson Microwave Anisotropy Probe (WMAP) \cite{Spergel:2003cb,Spergel:2006hy,Komatsu:2008hk,Komatsu:2010fb} implies that it is dominated by acoustic peaks arising from gravity-driven sound waves in the photon-baryon fluid. And then the positions of these acoustic peaks are shifted by cosmic expansion. Thus the positions and amplitudes of acoustic peaks contain important cosmic information. 

Baryons were strongly coupled to photons before the decoupling epoch. After recombination ($z_r\simeq1100$) baryons got free from Compton drag of photons and stayed at a fixed radius, i.e., sound horizon. It determines the first acoustic peak in CMB anisotropy that we can observe directly. Since the size of the sound horizon is predicted from theory, $l_s\sim H^{-1}(z_r)$, this sound horizon serves as a {\it standard ruler} by measuring the angular scale of the first acoustic peak. It is analogous to how SN Ia can be used as a standard candle but is completely independent from the SN Ia technique.

Galaxy clusters also leave an imprint on the CMB, the so-called Sunyaev-Zel'dovich (SZ) effect \cite{1970Ap&SS...7....3S}. The SZ effect stands for the distortion of the CMB spectrum through inverse Compton scatterings due to the collisions of the CMB photons and the high energy electrons in galaxy clusters. Measuring this distortion of the CMB, we can estimate masses of the clusters. Since the SZ effect is a scattering effect, its magnitude is redshift-independent; very distant clusters are as easy to detect as nearby clusters. When combined with accurate redshifts and mass estimates for the clusters, e.g. X-ray observation, the SZ effect plays a role of a standard ruler \cite{Cooray:2001av} and can be used as a distance indicator. This is the same method as comparing the absolute magnitude with the apparent one. The SZ effect may also provide interesting constraints on the dark energy.

The combination of CMB and SN Ia observations indicates that our Universe is spatially flat and the matter component is about one quarter of the critical density, $\Omega_m\simeq0.3$. 
This implies the need for dark energy as a missing component $\Omega_\Lambda=1-\Omega_m\simeq0.7$, which is indeed consistent with the value of $\Omega_\Lambda\simeq0.7$ suggested by SN Ia surveys \cite{Riess:1998cb,Perlmutter:1998np}.

Dark energy also influences on large angle anisotropies of CMB, the so-called integrated-Sachs-Wolfe (ISW) effect \cite{Sachs:1967er}. Variation of the gravitational potential during the epoch of cosmic acceleration leads to differential gravitational redshifts of photons and large scale perturbations. Analyses of spatial correlation of the large scale structure 
(LSS) of galaxies indicate that the Universe is not Einstein-de Sitter Universe ($\Omega_m =1$).

\begin{figure}[htb]
\label{observation}
  \centering
  \includegraphics[scale=0.9]{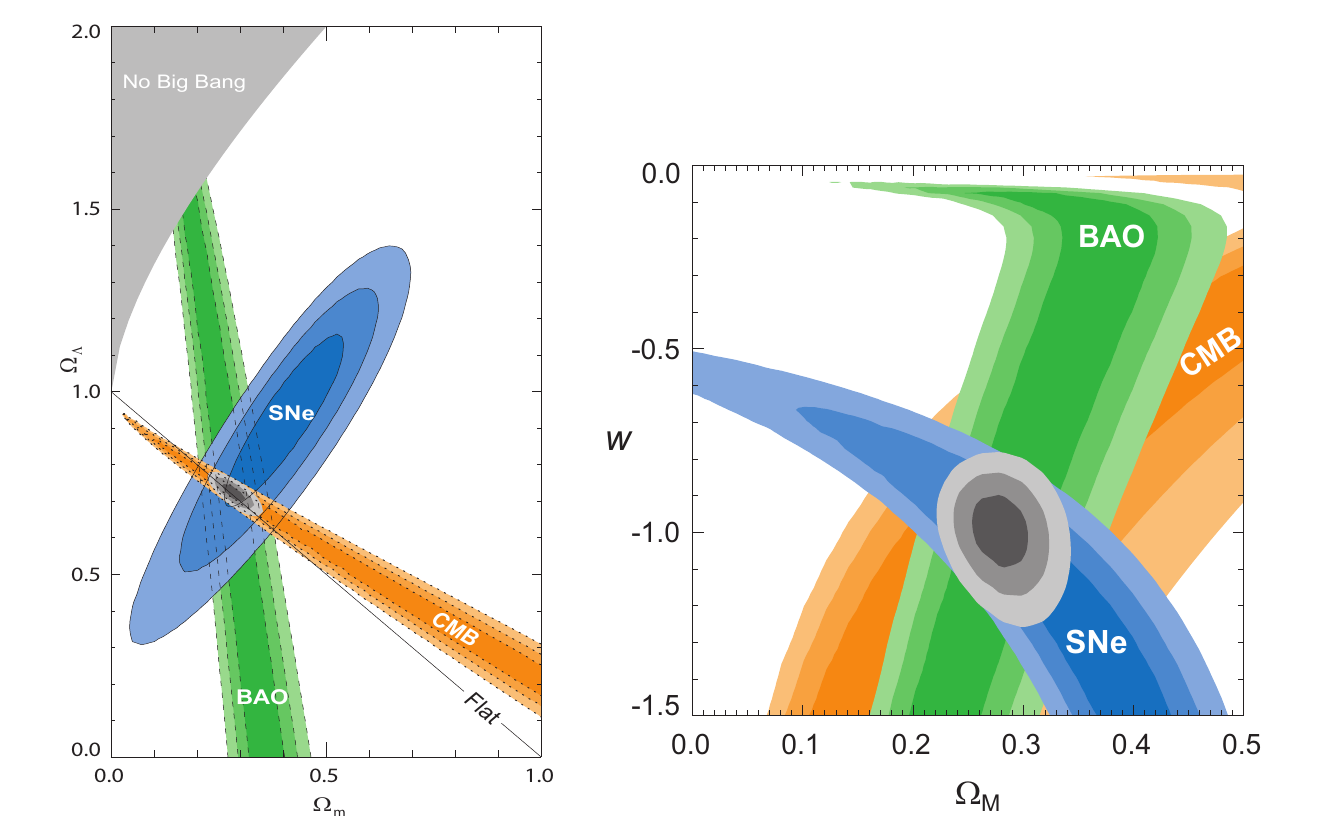}
  \caption[observation]{There is strong evidence for the existence of dark energy. Plotted are $\Omega_m$ - $\Omega_\Lambda$ (left panel) and  $\Omega_m$ - $w$ (right panel) confidence regions constrained from the observations of SN Ia, CMB and BAO. From Ref. \cite{Kowalski:2008ez}.}
\end{figure}

\subsubsection*{Baryon Acoustic Oscillations (BAO)}

Studying BAO is also a significant investigation for dark energy. As mentioned above, after decoupling baryons stay at the distance of the sound horizon whereas dark matter stays at the center of overdensity. They attract matter and eventually form galaxies. Therefore it is expected that a number of galaxies are separated by the sound horizon. We call it BAO signal. It means that by observing the large scale structure (LSS) of galaxies, we can measure the sound horizon scales and compare them with theoretical predictions.

To analyze the BAO signal, we do not need to resolve galaxy magnitudes or galaxy images. The only thing to be determined is their 3-dimensional position. Thus in that sense, BAO observations are safer from astronomical uncertainties than other dark energy probes.

The Sloan Digital Sky Survey (SDSS) catalog provides us a picture of the distribution of galaxies up to $z=0.47$ \cite{Eisenstein:2005su} and search for a BAO signal. To analyze this picture, we use a two-point correlation function of galaxies; it gives the probability that one galaxy will be found within a given distance of another. The two-point correlation function indeed possesses a bump, BAO signal, at a scale $100 h^{-1}$ Mpc for $0.16<z<0.47$. Both CMB and BAO signals indicate that the sound horizon today is about  $150$ Mpc.

Figure \ref{observation} shows us the best fit of cosmological parameters according to three independent observations. 

\subsubsection*{Weak gravitational lensing}

 The gravitational bending of light by structure distorts the images of galaxies. Studying this gravitational lensing effect, we can map dark matter and its clustering which shows us the influence of dark energy in the growth of large-scale structure. 
 
In rare cases we observe {\it strong lensing} which is deflection of light by massive structures like clusters and causes multiple images of the same background galaxy. More often we observe {\it weak lensing} which distorts the shapes, sizes and brightness of galaxies. Since we know the shape better than the original size and brightness of galaxy, many studies focus mainly on the change on shapes which is called {\it cosmic shear}. Weak lensing normally reads only $1\%$ change of galaxy shape. Therefore, a large number of galaxies are required to detect cosmic shear signal. The shear correlation function can be directly compared with the measurements \cite{Crittenden:2000au}.

There are dark energy probes other than ones raised above, including cosmic age tests from globular cluster and the white dwarfs \cite{Feng:2004ad}, galaxy clusters \cite{Allen:2011zs,Wang:1998gt,Haiman:2000bw}, Hubble parameter measurements \cite{Yu:2010pc,Freedman:2010xv} and gamma ray bursts \cite{Wang:2008vja,Tsutsui:2008cu,Basilakos:2008tp}. 
\subsection{Dark energy projects}

There are many observational projects to detect dark energy using the four dark energy probes: SN Ia, CMB, BAO and Weak Lensing (WL). Among them the Dark Energy Task Force (DETF) report \cite{Albrecht:2006um} classifies the dark energy projects into four stages: Stage I--completed projects that already released data, Stage II--on-going projects, Stage III--intermediate-scale, near-future projects, and Stage IV--large-scale, longer-term future projects. Table \ref{projects} shows us the dark energy projects classified by probes and stages.

\begin{table}[htb]
\begin{center}
    \begin{tabular}{ | c | p{4.3cm} | p{2.6cm} |  p{2.5cm}| p{2.5cm} |}
    \hline
    Probes & SN Ia & CMB & BAO & WL \\ \hline
    Stage I & Higher-Z Team \cite{Riess:2004nr,Riess:2006fw}, SNLS \cite{Astier:2005qq,Regnault:2009kd,Guy:2010bc}, ESSENCE \cite{Miknaitis:2007jd,WoodVasey:2007jb}, NSF \cite{Scalzo:2010xd}, CSP \cite{Folatelli:2009nm,Contreras:2009nt}, LOSS \cite{Leaman:2010kb,Li:2010kc,Li:2010kd}, SDSS \cite{Holtzman:2008zz,Kessler:2009ys}, SCP \cite{Kowalski:2008ez,Amanullah:2010vv,Suzuki:2011hu}, CfA  \cite{Hicken:2009dk,Hicken:2009df}, Palomar QUEST Survey \cite{Djorgovski:2008qg}&COBE \cite{Smoot:1992td}, TOCO \cite{Miller:1999qz}, BOOMERang \cite{deBernardis:2000gy}, Maxima \cite{Hanany:2000qf}, WMAP \cite{Spergel:2003cb,Spergel:2006hy,Komatsu:2008hk,Komatsu:2010fb} & 2dFGRS \cite{Colless:2003wz}, SDSS \cite{York:2000gk}, 6dFGRS \cite{Jones:2004zy}, WiggleZ \cite{Blake:2011wn}& CFHTLS \cite{Hoekstra:2005cs,Fu:2007qq}\\ \hline
    Stage II & Pan-STARRS1 \cite{Kaiser:2002zz}, HST \cite{Barbary:2010bv}, KAIT \cite{kait:website} & Planck \cite{Ade:2011ah,Planck:2011aj,Planck:2011ai}, SPT \cite{Ruhl:2004kv,Schaffer:2011mz}, ACT \cite{Kosowsky:2004sw} & SDSS II \cite{Tegmark:2006az}, SDSS III \cite{Ahn:2012fh}, BOSS \cite{Anderson:2012sa,Dawson:2012va}, LAMOST \cite{Deng:2012nv}, WEAVE \cite{Balcells:2010ck}&Pan-STARRS1, DLS \cite{2002SPIE.4836...73W,2009ApJ...702..603A}, KIDS \cite{deJong:2012zb} \\ \hline
    Stage III & DES \cite{Lin:2006ym}, Pan-STARRS4, ALPACA\cite{Corasaniti:2005kb},ODI \cite{odi:website}&ALPACA, CCAT \cite{Radford:2007hj}&DES, HETDEX \cite{Hill:2008mv}, BigBOSS \cite{Schlegel:2009uw}, ALPACA, SuMIRe \cite{sumire:website}&DES, Pan-STARRS4, ALPACA, ODI\\ \hline
    Stage IV &LSST \cite{Ivezic:2008fe}, WFIRST \cite{Green:2011zi}&EPIC \cite{Bock:2008ww,Bock:2009xw}, LiteBIRD \cite{Hazumi:2011zz, bird:website}, B-Pol \cite{deBernardis:2008bf}&LSST, SKA \cite{TorresRodriguez:2007mk}, WFIRST, Euclid \cite{Laureijs:2011mu} &LSST, SKA, WFIRST, Euclid\\
    \hline
    
    \end{tabular}
    \end{center}
\caption{Dark energy projects. Classification is taken from ref.~\cite{Albrecht:2006um}. Note that the DETF report was published in 2006, and thus many Stage II projects are now shifted to Stage I.}\label{projects}
\end{table}

We should mention that there exist many other projects planned to survey dark energy.
Those tremendous efforts may also help us to understand the origin and nature of dark energy.


\section{Cosmological constant}\label{sec:cc}

For explaining the observed accelerated expansion of the Universe, the simplest solution was to borrow Einstein's idea of vacuum energy, namely cosmological constant \cite{Carroll:2000fy,Peebles:2002gy}.

Einstein was seeking statistic solutions ($\dot{a}=0$), so he proposed a modification of his equation. Einstein's equation with the constant $\Lambda$ is given by
\begin{equation}
 R_{\mu\nu}-\frac{1}{2}g_{\mu\nu}R-\Lambda g_{\mu\nu}=8\pi GT_{\mu\nu}.
\end{equation}
With this modification, the Friedmann equations become
\begin{equation}
\frac{\dot{a}^2}{a^2}+\frac{k}{a^2}=\frac{8\pi G}{3}\rho+\frac{\Lambda}{3}
\end{equation}
and
\begin{equation}
\frac{\ddot{a}}{a}=-\frac{4\pi G}{3}(\rho+3p)+\frac{\Lambda}{3}.
\end{equation}
The corresponding action is 
\begin{equation}
\textsl{S}=-\frac{1}{16\pi G}\int d^4x\sqrt{-g}(R+2\Lambda)+\textsl{S}_M,
\end{equation}
where $g$ is the determinant of the metric tensor $g_{\mu\nu}$ and $R$ is the Ricci scalar.  And $\textsl{S}_M$ denotes the matter action. As we see here, the cosmological constant is a constant term in the Lagrange density. For the cosmological constant, $p_\Lambda=-\rho_\Lambda=-\frac{\Lambda}{8\pi G}=const.$, i.e., $w_\Lambda=-1$.

The small positive cosmological constant has been supported by a number of observations. Indeed the cosmological constant is a perfect fit to the dark energy data, even if we cannot explain it. There are two cosmological constant problems \cite{Weinberg:1988cp,Weinberg:2000yb,Peebles:2002gy}. The first one is why the vacuum energy is so small or does not gravitate at all, and the second one is why it is comparable to the present mass density.

\subsection{Fine-tuning problem}
 If the origin of $\Lambda$ is a vacuum energy, there is a serious problem of its energy scale.
The vacuum means a state of minimum energy. For example, for a harmonic oscillating particle the potential is of the form $V(x)=\frac{1}{2}\omega^2 x^2$. In this case, the vacuum state is given by the particle sitting motionless in the minimum of the potential, i.e., $x=0$. But quantum mechanically we cannot obtain position and momentum of the particle at the same time because of the uncertainty principle. Instead, we know that the lowest energy state has an energy $E_0=\frac{1}{2}\hbar\omega$. 

We could consider a quantum field as an infinite set of harmonic oscillators, where the minimum energy of such a field should be also infinite. However, if we trust our theory only up to a certain cutoff, like the Planck scale $M_{pl}$, from dimensional consideration we would get the form
\begin{equation}
\rho_\Lambda\sim\hbar M_{pl}^4.
\end{equation}
Indeed, we measure the vacuum fluctuations by the Casimir effect in the laboratory.

For the Planck scale  $M_{pl}=(8\pi G)^{-1/2}\sim10^{18}GeV$, we expect
\begin{equation}
\rho_\Lambda \sim (10^{18}GeV)^4 \sim 2\times10^{110}erg/cm^3.
\end{equation}
However, most of cosmological observations imply
\begin{equation}
\rho_\Lambda^{obs}\leq(10^{-12}GeV)^4\sim2\times10^{-10}erg/cm^3.
\end{equation}
There is 120 orders of magnitude difference between the theoretical expectation and the observational value. This discrepancy has been called  \textquoteleft the worst theoretical prediction in the history of physics!'.

Why is the vacuum energy so small? Does it somehow cancel out exactly a factor of $10^{120}$? These are major outstanding issues in physics and cosmology. Some supersymmetric theories predict a cosmological constant that is exactly zero. In supersymmetric theories, the number of fermionic and bosonic degrees of freedom are equal. The energy of the vacuum fluctuations per degree of freedom is the same in magnitude but opposite in sign for fermions and bosons of the same mass. Therefore the fermion and boson contributions cancel each other and the total vacuum energy density vanishes \cite{Mukhanov:2005sc}. But it is not helpful  to solve the problem because supersymmetry has to be broken today, as it is not observed in nature.

If supersymmetry is broken, supersymmetric partners can have different masses of order $\Lambda^{4}_{SUSY}$, where $\Lambda_{SUSY}$ is the supersymmetry breaking scale. Assuming $\Lambda_{SUSY}\sim 1TeV$, the vacuum energy density becomes $\rho_{\Lambda}\sim(10^{3} GeV)^{4}$ which is still 60 orders of magnitude larger than the observational limit.

\subsection{Coincidence problem}

The second cosmological constant problem is  why $\rho_\Lambda$ is not only small but also of the same order of magnitude as the present mass density of the Universe. In other words, why does cosmic acceleration happen to begin right now and not at some point in the past or future? 

If the vacuum energy were big and dominant from the earlier epoch, there would be no chance to form structures in the Universe, like galaxies, stars, planets and us, intelligent lives.
In other words, observers will only observe the states which are allowed for observers. This consideration is the so-called anthropic principle. According to the anthropic principle, the Universe may not be determined directly by one specific process, but there are different expanding regions at different times and space or of different terms in the wave function of the Universe. Therefore, the value of $\rho_\Lambda$ that we measure should be just suitable for the evolution of intelligent lives.

This anthropic consideration seems to give us the explanation for both of the two cosmological constant problems, why it is small and why acceleration starts now. Actually, we sometimes use the anthropic principle as a common sense. Why does the Earth lie exactly on the habitability zone, the narrow range of distance from the Sun at which the surface temperature allows the existence of liquid water? The answer could be that, if the Earth was not there, we would not be here.

But still, many scientists are not satisfied with this explanation and are disappointed by that they could not explain directly the observed Universe from first principles.

We are living $in$ our Universe, so we cannot verify whether the anthropic principle solves the cosmological constant problem or not. The anthropic explanation of the value $\rho_\Lambda$ makes sense only if there is a multiverse with a lot of big bangs and different values of $\rho_\Lambda$.

\section{Quintessence}\label{sec:quintessence}

The Einstein equation $G_{\mu\nu}=8\pi GT_{\mu\nu}$ determines the dynamics of the Universe. (See appendix (\ref{subsec:FLRW}) for basics.) From the cosmological constant problem we are motivated to find an alternative explanation of dark energy. By modifying the left hand side of the Einstein's equation, we get the modified gravity models. By modifying the right hand side, we get the modified matter models. The idea of modified matter models is that the energy momentum tensor $T_{\mu\nu}$ contains an exotic matter, which provides negative pressure. The following four sections are devoted to this possibility.

The first suggestion to solve the cosmological constant problems is quintessence \cite{Wetterich:1987fm,Ratra:1987rm,Caldwell:1997ii,Zlatev:1998tr}. The name quintessence means the fifth element, other than baryons,  dark matter, radiation and spatial curvature. This fifth element is the missing cosmic energy density component with negative pressure which we are searching for today.

 The basic idea of quintessence is that dark energy is in the form of a time varying scalar field which is slowly rolling down toward its potential minimum. 
 The full action including  quintessence is given by
 \begin{equation}
 \label{4.1}
\textsl{S}=\int d^4x\sqrt{-g}\left[-\frac{1}{16\pi G}R+\frac{1}{2}g^{\mu\nu}\partial_{\mu}\phi\partial_{\nu}\phi-V(\phi)\right]+\textsl{S}_M.
 \end{equation}
 The evolution of the scalar field is governed by 
\begin{equation}
\label{4.2}
\ddot{\phi}+3H\dot{\phi}+V'(\phi)=0,
\end{equation}
where $\phi$ is assumed to be spatially homogeneous.

 The energy density and the pressure of the scalar field are 
\begin{equation}
\rho_Q=\frac{1}{2}\dot{\phi}^2+V(\phi),  \indent  p_Q=\frac{1}{2}\dot{\phi}^2-V(\phi),
\end{equation}
where $\frac{1}{2}\dot{\phi}^2$ is the kinetic energy and $V(\phi)$ is the potential energy.

 We define the equation of state parameter $w$ as
\begin{equation}
\label{4.4}
w_Q=\frac{p_Q}{\rho_Q}=\frac{\frac{1}{2}\dot{\phi}^2-V(\phi)}{\frac{1}{2}\dot{\phi}^2+V(\phi)}
\end{equation}
which has a range of $-1<w<1$. In this range, we are interested in the negative pressure:  $-1<w\leq 0$. If the scalar field evolves very slowly so that the kinetic energy term is much smaller than the potential energy term, then $w$ is close to $-1$ and the scalar field behaves just like the cosmological constant.

\subsection{Thawing or freezing}
In the equation of motion (\ref{4.2}), the second term acts as a friction term. By varying the friction term, the quintessence model can be dynamically classified \cite{Caldwell:2005tm}.

\begin{itemize}
	
	\item Thawing model
	
	In the thawing model, the field at early times has been frozen by the Hubble friction term $3H\dot{\phi}$, and it acts as vacuum energy. When the expansion rate drops below $H<\sqrt{V''(\phi)}$, i.e. underdamped, then the field starts to roll down to the minimum and $w$ evolves away from -1. Examples of this model are   $V(\phi)=M^{4-\alpha}\phi^{\alpha}$ for $\alpha>0$, and $V(\phi)=M^4\exp(-\beta\phi/M_P)$ for $\beta<\sqrt{24\pi}$ \cite{Caldwell:2005tm}.
	
	\item Freezing model
	
	If the field is already rolling towards its potential minimum and slowing down, we have $H>\sqrt{V''(\phi)}$. In this case, the field is overdamped and approximately constant. Examples of this model are $V(\phi)=M^{4+\alpha}\phi^{-\alpha}$ and $V(\phi)=M^{4+\alpha}\phi^{-\alpha}\exp(\gamma\phi^2/M^2_P)$ for $\alpha>0$ \cite{Caldwell:2005tm}. In the first case the scalar field energy density tracks that of the dominant component (radiation, matter) at early times and eventually dominates at late times. Therefore, the tracker behavior partly solves the coincidence problem.	
\end{itemize}

\subsection{Tracker solution}\label{subsec:traker_sol}
In the tracker solution \cite{Zlatev:1998tr,Weinberg:2000yb}, the quintessence component tracks the background density for most of the history of the Universe, then only recently grows to dominate the energy density and drives the Universe into a period of accelerated expansion.

The simplest form of the tracker solution is obtained from

\begin{equation}
\label{4.5}
V(\phi)=M^{4+\alpha}\phi^{-\alpha},
\end{equation}
where $\alpha>0$ and the value of $M$ is fixed by the measured value of $\Omega_m$. 

\begin{figure}
  \begin{center}
  \input{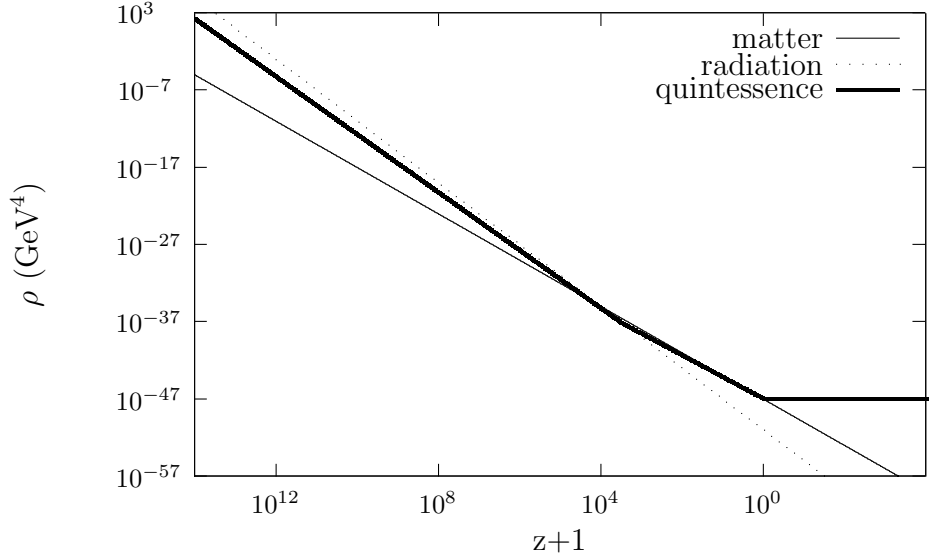}
  \caption[tracker]{A plot of energy density vs. redshift using the equation of state for a tracker solution. For an illustration, $\alpha=30$ in the Eq.(\ref{4.10}) is plotted. The dotted line is for radiation, the solid line is for matter, and the thick solid line is for quintessence. }
  \label{tracker}
  \end{center}
\end{figure}

\begin{itemize}
\item Early epoch (radiation and matter dominant epochs)

If the scalar field takes a value much less than the Planck mass in the beginning, its energy density is initially given by $ \rho_Q\ll\rho_M$, where the background energy density $\rho_M=\rho_m+\rho_{r}$. Using the equation of motion for the scalar field with the Eq.(\ref{4.5}), we have a solution of the field $\phi(t)$ which initially increases as \cite{Weinberg:2000yb}
 \begin{equation}
\phi(t) = \phi_0\left(\frac{t}{t_0}\right)^\frac{2}{2+\alpha},
 \end{equation}
 where we define a dimensionless quantity $\tilde{t}\equiv\frac{t}{t_0}$. Then,    $\rho_Q=\frac{1}{2}\dot{\phi}^2+V(\phi)$ decreases as $t^\frac{-2\alpha}{2+\alpha}$ :
 \begin{equation}
 \label{4.7}
 \rho_Q=\frac{1}{2}\dot{\phi_0}^{2}\cdot \tilde{t}^{\frac{-2\alpha}{2+\alpha}}+M^{4+\alpha} \tilde{t}^{\frac{-2\alpha}{2+\alpha}} \indent \propto \indent  t^{\frac{-2\alpha}{2+\alpha}}.
  \end{equation}
 Now the energy density of matter and radiation  $\rho_M$ is decreasing faster than the Eq.(\ref{4.7}) since
  \begin{equation}
  \rho_M=\rho_m+\rho_r \indent  = \indent \rho_{m,0} \left(\frac{a_0}{a_m}\right)^{3}+\rho_{r,0}\left(\frac{a_0}{a_r}\right)^{4}\indent  \propto\indent  t^{-2},
   \end{equation}
where the scale factor in the matter dominated Universe behaves as $a_m(t)\propto t^{2/3}$ and in the radiation dominated Universe as $a_r(t)\propto t^{1/2}$.
Therefore, the matter dominant epoch cannot last forever since $\rho_Q$ dominates eventually.

\quad Does the tracker solution really track the background? To derive the relation between the equation of state of the field $w_Q$ and 
that of the background $w_M$, we use the known properties,  $\rho_Q\propto a^{-3(1+w_Q)}$ and  $a\propto t^{\frac{2}{3(1+w_M)}}$ :
\begin{equation}
\label{4.9}
 \rho_Q \propto  t^{\frac{-2\alpha}{2+\alpha}}\indent \propto a^{-3(1+w_Q)}\indent \propto t^{\frac{-2(1+w_Q)}{1+w_M}},
\end{equation}
where $Q$ denotes quintessence and $M$ does the background. Comparing exponents of $t$ in the Eq.(\ref{4.9}), we finally get the equation of state parameter for the tracker solution as \cite{Zlatev:1998tr}
\begin{equation}
\label{4.10}
w_Q\approx\frac{\frac{\alpha}{2}w_M-1}{1+\frac{\alpha}{2}}
\end{equation}
which is valid so long as $\rho_M\gg \rho_Q$. This solution is plotted in Fig.\ref{tracker}.

\quad  For $\alpha\gg1$, we show that $w_Q\approx w_M$ , i.e.,  $w_Q\approx \frac{1}{3}$ during the radiation dominated epoch and $w_Q\approx0$ during the matter dominated epoch. In general, $w_Q$ depends on both the effective potential Eq.(\ref{4.5}) and $w_M$. The effect of the background comes from the Eq.(\ref{4.2}), where $H$ carries the information of the background. Another remarkable feature is that $w_Q$ decreases to a negative value (negative pressure that we want) after the transition from the radiation dominated epoch to the matter dominated epoch regardless of the initial value of $w_Q$.

\item{Later epoch (quintessence dominant epoch)}

In later epoch,  $\rho_Q$ becomes relatively larger and at some time catches up with $\rho_M$, then $\rho_Q$ decreases more slowly as
\begin{equation}
 \rho_Q\propto t^{-\frac{2\alpha}{4+\alpha}}.
 \end{equation}
 This is because the field value $\phi(t)$ increases as
 \begin{equation}
  \phi(t)\propto t^{\frac{2}{4+\alpha}}.
  \end{equation}
   The expansion rate $H$ now goes as $H\propto\sqrt{\rho_Q}\propto\sqrt{V(\phi)}\propto t^{-\frac{\alpha}{4+\alpha}}$, so the scale factor $a(t)$ grows almost exponentially
  \begin{equation}
   \ln{a(t)} \propto t^\frac{4}{4+\alpha}.
   \end{equation}
    Thus the quintessence field drives the late time accelerated expansion of the Universe. In this solution the transition from $\rho_M$ dominance to $\rho_Q$ dominance is supposed to take place near the present time so that both $\rho_M$ and $\rho_Q$ are now contributing to the cosmic expansion.
    
    \quad One could also confirm directly that $w_Q\approx -1$ for the later epoch by
    \begin{equation}
    w_Q=\frac{\frac{1}{2}\dot{\phi}^2-V(\phi)}{\frac{1}{2}\dot{\phi}^2+V(\phi)}=\frac{\frac{1}{2}\left(\frac{2}{4+\alpha}\right)^2\left(\frac{\phi_0}{t_0}\right)^2 \tilde{t}^{\frac{-4-2\alpha}{4+\alpha}}-M^{4+\alpha}\phi^{-\alpha}_0\tilde{t}^{\frac{-2\alpha}{4+\alpha}}}{\frac{1}{2}\left(\frac{2}{4+\alpha}\right)^2\left(\frac{\phi_0}{t_0}\right)^2 \tilde{t}^{\frac{-4-2\alpha}{4+\alpha}}+M^{4+\alpha}\phi^{-\alpha}_0\tilde{t}^{\frac{-2\alpha}{4+\alpha}}}\indent \approx -1,
    \end{equation}
       where the potential term $V$ becomes more important than the kinetic term as $t\rightarrow \infty$.
\end{itemize}

As the name $ tracker$ $solution$ says, the equation of state $w_Q$ tracks that of the background $w_M$. When radiation dominates ($w_M=\frac{1}{3}$), $w_Q\leq\frac{1}{3}$ and $\rho_Q$ decreases less rapidly than $\rho_r$. When matter dominates ($w_M=0$), $w_Q<0$ and  $\rho_Q$ decreases less rapidly than $\rho_m$. Finally $\rho_Q$ becomes the dominant component and $w_Q\rightarrow -1$, as the Universe enters to the accelerating phase. See Fig.\ref{tracker}.
At the transitions from radiation to matter domination and from matter to quintessence domination, $w_Q$ changes almost instantaneously.

The big advantage of the tracker solution is that, thanks to the existence of a crossover, the model does not depend sensitively on the initial conditions for the field. However different the initial conditions were, the scalar field keeps rolling down slowly and at the end their solutions behave similarly. This is the attractor behavior of the tracker solution. Actually, the initial ratio of the tracker field $\rho_Q $ to the matter density $\rho_m$ can vary by nearly 100 orders of magnitude and does not affect the cosmic history \cite{Zlatev:1998tr}. So the quintessence model says that the dark energy density is so small because the Universe is old.

Does it mean that the two cosmological constant problems are solved? The answer is no. To make the value of $\rho_Q$ at which  $\rho_Q\approx\rho_M$ close to the present critical density $\rho_{c0}$, we need again fine-tuning of potential energy.

There is a rough estimate of which value the $M$ should have \cite{Weinberg:2000yb}.
If we assume that the value of the field $\phi(t)$ at this crossover is of the order of the Planck mass, $M_{pl}=(8\pi G)^{-1/2}$, the scalar field evolves very slowly so that $V(\phi)\approx\rho_\phi\approx\rho_{c0}=\frac{3H_0^{2}}{8\pi G} $. From the Eq.(\ref{4.5})  we need
\begin{equation}
M^{4+\alpha}\approx (8\pi G)^{-\alpha/2}\rho_{c0}\approx (8\pi G)^{-1-\alpha/2}H_0^2.
\end{equation}
However, the quintessence theory gives no explanation why $M$ should have this value.

Besides, in order for the field to be slowly rolling today, we require $\sqrt{V''(\phi)}\sim H_0$ that corresponds to an effective mass $m_\phi$ of $\phi$. Thus we must have

\begin{equation}
m_\phi \sim H_0\sim10^{-33}eV,
\end{equation}
 which is a very small number in particle physics.

\section{K-essence}\label{sec:kessence}

The quintessence model assumes the canonical kinetic energy term $\frac{1}{2}\partial^{\mu}\phi\partial_{\mu}\phi$ and the potential energy term $V(\phi)$ in the action. Modifying this canonical kinetic energy term, the non-canonical (non-linear) kinetic energy of the scalar field can drive the negative pressure without the help of potential terms \cite{ArmendarizPicon:1999rj}. The non-linear kinetic energy terms are thought to be small and usually ignored because the Hubble expansion damps the kinetic energy density over time. But what happens if there is a dynamical attractor solution which forces the non-linear terms to remain non-negligible? This is the main idea of the k-essence \cite{ArmendarizPicon:2000dh,ArmendarizPicon:2000ah}.

The full action including a k-essence term is given by
\begin{equation}
\label{4.19}
\textsl{S}=\int d^4x\sqrt{-g}\left[-\frac{1}{16\pi G}R+p(\phi,X)\right]+\textsl{S}_M,
\end{equation}
where $X$ is the canonical kinetic energy of the field,
\begin{equation}
\label{4.20}
X\equiv \frac{1}{2}(\nabla\phi)^{2},
\end{equation}
and the Lagrangian $p(\phi,X)$  plays a role of the pressure $p_K$. Here we consider a model with
\begin{equation}
p_K=p(\phi,X)=\tilde{p}(X)/\phi^{2},
\end{equation}
which has the desired property for dark energy.

For small $X$,  $\tilde{p}(X)$ could be expanded as  $\tilde{p}(X)= const. +X+\mathcal{O}(X^{2})$. If we ignore the non-linear term $\mathcal{O}(X^{2})$ and take an additional potential, then we come back to the quintessence model. The scalar field for which these higher order kinetic energy terms play an essential role is k-essence.

The energy density of the k-field is
\begin{equation}
\label{4.22}
\rho_K=(2X\tilde{p}_{,X}-\tilde{p})/\phi^{2}\equiv \tilde{\rho}/\phi^{2}
\end{equation}
so that the equation of state parameter for the k-field is
\begin{equation}
w_K\equiv \frac{p_K}{\rho_K}=\frac{\tilde{p}}{\tilde{\rho}}=\frac{\tilde{p}}{2X\tilde{p}_{,X}-\tilde{p}}
\end{equation}
where $_{,X}$ means derivative with respect to $X$. If the Lagrangian $p$ satisfies the condition $Xp_{,X}\ll p$ for some range of $X$ and $\phi$, then the equation of state is $p\approx -\rho$ so that we have an accelerated expansion solution. We shell find the form of  $\tilde{p}(X)$ to satisfy this requirment.

The effective speed of sound $c_s$ of k-essence is defined by \cite{Garriga:1999vw}
\begin{equation}
\label{4.24}
c_s^{2}=\frac{p_{,X}}{\rho_{,X}}=\frac{\tilde{p}_{,X}}{\tilde{\rho}_{,X}}.
\end{equation}

From observations we know that our Universe is almost flat, so we ignore the curvature term in the Friedmann equation :
\begin{equation}
H^{2}\equiv \dot{N}^{2}=\frac{8\pi G}{3}(\rho_M+\rho_K), \indent N\equiv \ln a.
\end{equation}

The energy conservation equations for the k-essence ($i\equiv K$) and the background ($i\equiv M$) components are
\begin{equation}
\label{4.26}
\frac{d\rho_i}{dN}=-3\rho_i(1+w_i),
\end{equation}
where $w_i$ is the equation of state for the corresponding matter (dust and radiation) or k-essence.
Plugging the energy density (\ref{4.22}) in the conservation equation (\ref{4.26}) and considering a homogeneous field $\phi$, we get 
\begin{equation}
\label{4.27}
\frac{dX}{dN}=-\frac{\tilde{\rho}}{\tilde{\rho},_X}\left[3(1+w_K)-2\phi^{-1}\frac{\sqrt{2X}}{H}\right].
\end{equation}

Now we consider the conditions for the stability of the field. We require positive energy density,
\begin{equation}
\tilde{\rho}=2X\tilde{p}_{,X}-\tilde{p}>0,
\end{equation}
 $c_s^{2}>0$ and the function $\tilde{\rho}(X)$ to increase monotonically with $X$,
\begin{equation}
\tilde{\rho}_{,X}=2X\tilde{p}_{,XX}+\tilde{p}_{,X}>0.
\end{equation}

It is convenient to use a new variable $y\equiv 1/\sqrt{X}$ and to rewrite the pressure and the energy density of the k-field as 
\begin{equation}
p_K\equiv g(y)/(\phi^{2}y) , \indent \rho_K\equiv -g^{\prime}/\phi^{2}.
\end{equation}
In this case the equation of state is
\begin{equation}
w_K=-g/(y g^{\prime})
\end{equation}
and the sound speed is
\begin{equation}
c_s^{2}=\frac{p_K^{\prime}}{\rho_K^{\prime}}=\frac{g-g^{\prime}y}{g^{\prime\prime}y^{2}},
\end{equation}
where prime denotes derivative with respect to $y$.

Using this new variable, the stability conditions can be expressed as
\begin{equation}
\tilde{\rho}=-g^{\prime}>0
\end{equation}
and
\begin{equation}
\tilde{\rho}_{,X}=\frac{1}{2}y^{3}g^{\prime\prime}>0.
\end{equation}
Equally  $g^{\prime}<0$ and $g^{\prime\prime}>0$ so that $g$ is a decreasing convex function of $y$. See the Figure 4.2.

\begin{figure}[htb]
\label{kessence}
  \centering
  \includegraphics[scale=0.8]{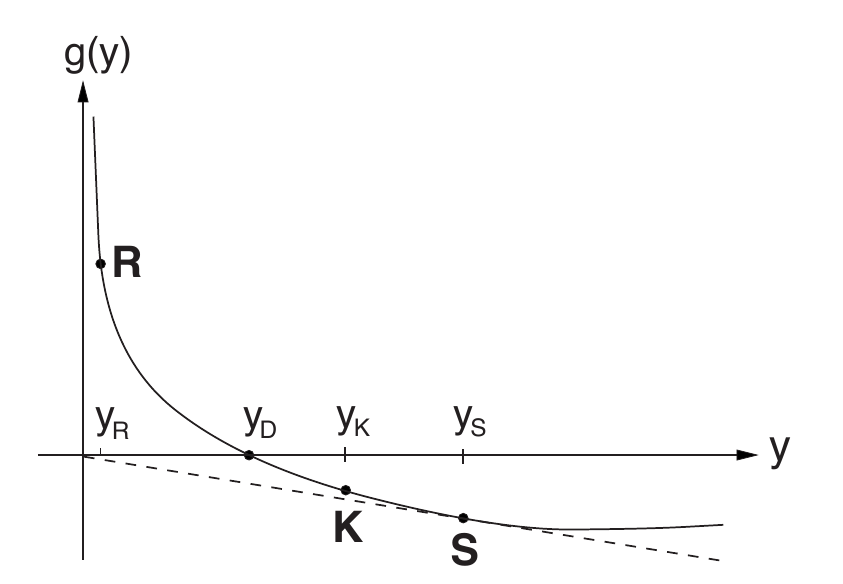}
  \caption[Kurzform f"ur das Abbildungsverzeichnis]{A plot of $g(y)$. The points indicate the different attractor solutions. $\mathbf{ R}$ is for the attractor solution during the radiation dominated epoch, $\mathbf{ S}$ is the de Sitter attractor at the onset of matter dominated epoch and $\mathbf{ K}$ is  the attractor as k-essence dominates. The tangent to the curve at the de Sitter attractor goes through the origin. From Ref.(\cite{ArmendarizPicon:2000dh})}
\end{figure}

Rewriting the equation (\ref{4.27}) in terms of the new variable, we get the equation of motion for the k-field,
\begin{equation}
\label{4.35}
\frac{dy}{dN}=\frac{3}{2}\frac{(w_K(y)-1)}{r^{\prime}(y)}\left[r(y)-\sqrt{\frac{\rho_K}{\rho_{tot}}}\right],
\end{equation}
where 
\begin{equation}
\label{4.36}
r(y)\equiv \left(-\frac{9}{8}g^{\prime}\right)^{1/2}y(1+w_K)=\frac{3}{2\sqrt{2}}\frac{(g-g^{\prime}y)}{\sqrt{-g^{\prime}}}.
\end{equation}
From (\ref{4.26}) we get 
\begin{equation}
\label{4.37}
\frac{d(\rho_K/\rho_{tot})}{dN}=3\frac{\rho_K}{\rho_{tot}}\left(1-\frac{\rho_K}{\rho_{tot}}\right)(w_M-w_K(y)).
\end{equation}
These equations of motion describe the dynamics of the k-essence model.

\subsection{Attractor solutions}
If every point that is close to some point $A$ is attracted to $A$, we call the point $A$ an $attractor$. The attractor solutions for k-essence are classified into two types. The first one is tracker solution in which k-essence mimics the equation of state of the background component in the Universe. In the second case k-essence is attracted to an equation of state which is different from matter or radiation.
 
We want to have a tracker solution $y(N)$ which satisfies $w_K(y(N))=w_M=const.$ 
\begin{equation}
\label{4.38}
w_K(y_{tr})=-\frac{g}{yg^{\prime}}|_{y=y_{tr}}=w_M
\end{equation}
and $\rho_K/\rho_{tot}$ is fixed so that the solution satisfies $y(N)=y_{tr}=const.$.  Hence $y_{tr}$ changes depending on the epoch and $w_M$. The point $y_{tr}$ that satisfies those conditions is the so-called attractor solution. Plugging this solution, $dy/dN=0$ and $d(\rho_K/\rho_{tot})/dN=0$, in the equations of motion for k-essence (\ref{4.35}),(\ref{4.37}) we get a simple relation
\begin{equation}
\label{4.39}
r^{2}(y_{tr})=\frac{\rho_K}{\rho_{tot}}<1,
\end{equation}
where we considered $\rho_K<\rho_{tot}$.
I.e., in the range of $r(y)>1$ there is no attractor solution.

 Let us consider four possible attractors for k-essence in the different epochs. We consider positions of attractor solutions, characterize attractor solutions using the Eq.(\ref{4.38}) and check existence of attractor solutions using the Eq.(\ref{4.39}).

\begin{itemize}
  \item Radiation tracker
  
  Radiation has positive pressure and the k-essence pressure $p_K$ is proportional to $g(y)$.  Thus the radiation attractor should be located in the region $g(y)>0$, equally in the range $y<y_D$, where $y_D$ is the point which satisfies $g(y_D)=0$. See Figure 4.2. 
  The radiation attractor satisfies $w_K(y_R)=w_R=\frac{1}{3}$. By substituting this in the equation (\ref{4.38}), we get
  \begin{equation}
y_R g^{\prime}(y_R)=-3g(y_R).
\end{equation}
  During the radiation dominated epoch the k-essence energy density is attracted on the radiation attractor and fixed by 
 \begin{equation}
 \left( \frac{\rho_K}{\rho_{tot}}\right)_R=r^{2}(y_R)=  -2g^{\prime}(y_R)y_R^{2},
\end{equation}
where the Eq. (\ref{4.36}) has been used.

The most likely range for $r^{2}(y_R)= \left(\frac{\rho_K}{\rho_{tot}}\right)_R$ is $10^{-1}\sim 10^{-2}$ \cite{ArmendarizPicon:2000ah}, so that the cosmic acceleration begins at the present epoch. If $r^{2}(y_R)$ is much smaller than $10^{-2}$, the k-essence energy density at the dust-radiation equality is so small that it would not have overtaken the matter density today. On the other hand, if $r^{2}(y_R)$ is much greater than $10^{-1}$,  the expansion rate in the early Universe would be so huge that it would spoil the predictions of the primordial nucleosynthesis \cite{Olive:1999ij}.

  \item Dust tracker
  
  The dust attractor  corresponds to zero pressure so that $w_K(y_D)=w_D=0$ and $g(y_D)=0$ at the  dust attractor $y=y_D$. Thus this attractor should locate at the point where $g(y)$ goes through zero. For the dust attractor, we have
  \begin{equation}
 \left( \frac{\rho_K}{\rho_{tot}}\right)_D=r^{2}(y_D)=  -\frac{9}{8}g^{\prime}(y_D)y_D^{2}.
\end{equation}
   As we discussed above, the dust attractor exists only if $r^{2}(y_D)<1$.
  
  \item De Sitter attractor
  
  Now we consider the radiation-dust equality phase and at that point the k-essence energy density is negligibly small, $\rho_K\ll\rho_M$. Then we have $r^{2}(y_S)=\left(\frac{\rho_K}{\rho_{tot}}\right)_S\rightarrow 0$ as $y(N)\rightarrow y_S$ in the equation (\ref{4.35}). Since $r\propto (1+w_K)$, we finally get $w_K(y_S)\approx-1$ and 
  \begin{equation}
g(y_S)=g^{\prime}(y_S)y_S,
\end{equation}
 where the tangent of $g(y)$ at $y_S$ passes through the origin. See Figure 4.2. Note that we can always find such a point $y_S$ for a decreasing convex function $g(y)$. Thus the de Sitter attractor (S-attractor) is a generic feature of the k-essence models.
 
 The assumption $\rho_K\ll\rho_M$ at the dust-radiation equality is also required to satisfy the nucleosynthesis constraints \cite{Olive:1999ij}. In this scenario, the k-essence approaches to the de Sitter attractor shortly after the onset of the matter dominate epoch and behaves like a cosmological constant, $w_K\rightarrow -1$. The k-essence energy density $\rho_K$ becomes very small and freezes at some fixed value. 
 
 Since the matter density $\rho_M$ decreases while $\rho_K$ remains constant, the k-essence density eventually overtakes the matter density. In this regime, the relation $\rho_K\ll\rho_M$ does not hold, the k-field moves to the K-attractor as described below.
 
  \item K-attractor
  
  When $\rho_K$ overtakes $\rho_M$ and dominates,  $\frac{\rho_K}{\rho_{tot}}\rightarrow 1$, then according to the equation (\ref{4.35}), it follows $r(y_K)=1$ as $y(N)\simeq y_K$. This solution describes a power-law expanding Universe as follows \cite{ArmendarizPicon:1999rj}:  substituting $r(y_K)=1$ in the definition of the $r(y)$ Eq.(\ref{4.36}), we get the relation
  \begin{equation}
1+w_K(y_K)=\frac{2\sqrt{2}}{3}\frac{1}{\sqrt{-g_K^{\prime}y_K^{2}}}=const.
\end{equation}
  and it follows 
  \begin{equation}
a\propto t^{\frac{2}{3(1+w_K)}} = t^{\sqrt{-g^{\prime}_K y_K^{2}/2}}.
\end{equation}
If it satisfies  $-g^{\prime}_K y_K^{2}/2>1$, then the solution describes power-law inflation.

Now let us consider the existence of the K-attractor. We restrict the range of K-attractor to  $y_D<y_K<y_S$ so that the k-essence dominant Universe does not have positive pressure. If $r(y_D)>1$, i.e., there is no dust attractor but exists S-attractor [$r(y_S)=0$], then the K-attractor [$r(y_K)=1$] must exist between $y_D<y_K<y_S$ since $r(y)$ is a continuous function. In this interval $y>y_D$, it has negative pressure [$g(y_K)<0$] and induces the power-law cosmic acceleration. On the other hand if  $r(y_D)<1$, i.e., there is a dust attractor, then since $r^{\prime}(y)<0$ for $y>y_D$, where
\begin{equation}
r^{\prime}=\frac{3}{4\sqrt{2}}\frac{g^{\prime\prime}y}{\sqrt{-g^{\prime}}}(w_K-1),
\end{equation}
 there is no point $y=y_K>y_D$ where $r(y)=1$. It means, there is no K-attractor at $y_D<y<y_S$. 
  \end{itemize}

As we considered above, there are two possible scenarios; one is without dust attractor  (R $\rightarrow$ S $\rightarrow$ K-attractor) and the other one is with dust attractor   (R $\rightarrow$ S $\rightarrow$ D-attractor). According to the first scenario, k-essence is attracted to $y=y_R$ in the radiation dominated epoch. At matter dominated epoch, $\rho_K$ drops sharply by several orders of magnitude and k-essence skips the point $y=y_D$. As $y\approx y_S$, $\rho_K$ freezes and overtakes $\rho_m$. And then $y$ relaxes towards $y_K$. In this scenario, our current Universe lies on the transition from $y_S$ to $y_K$.

For the second scenario, with dust attractor [$r^{2}(y_D)<1$], we consider again two situations. If $r^{2}(y_D)=\left(\frac{\rho_K}{\rho_{tot}}\right)_D\ll1$, the k-essence could neither dominate today nor cause the cosmic acceleration. In the case $r^{2}(y_D)=\left(\frac{\rho_K}{\rho_{tot}}\right)_D\rightarrow 1$, the expansion of the Universe would be accelerated before the k-field reaches the dust attractor. In this case, the k-essence approaches first the S-attractor, freezes for a finite time, is attracted towards the dust attractor, and the Universe decelerates its expansion. This scenario is called $late$ $dust$ $tracker$ because the dust attractor is reached long after the matter domination has begun. According to the late dust tracker scenario, the cosmic expansion returns to pressureless, unaccelerated in the long-term future.

\subsection{Comparing to Quintessence}
Our goal is to have a model which solves the cosmological constant problems. In that sense, the dynamical attractor behavior of both the quintessence and the k-essence model have a big advantage. The cosmic evolution in this model is insensitive to initial conditions because the k-field is attracted to the attractor solution wherever it started. Moreover it solves the coincidence problem by explaining why the cosmic acceleration is started at such a late stage shortly after the onset of the matter dominated phase. However, both of the quintessence and the k-essence do not solve vacuum energy problem.

Comparing to the tracker solution in the quintessence model, the quintessence field tracks the radiation and matter background, and needs a potential energy fine-tuning at the quintessence-matter crossover stage. But the k-essence field tracks only the radiation background (for no D-attractor scenario), and does not need a potential energy term thus it is free from fine-tuning that arose in quintessence.

For no D-attractor scenario, $w_K$ is increasing today from $-1$ towards its asymptotic value at K-attractor. The numerical value of the effective equation of state in this case is $w_{eff}=-0.84$ \cite{ArmendarizPicon:2000dh}. According to the quintessential tracker solution, the current Universe undergoes a phase from $w\approx0$ to $w=-1$ and the effective equation of state value is $w_{eff}\approx-0.75$ \cite{Steinhardt:1999nw}. The supernovae observation data are more consistent with the k-essence model.

On the other hand, during the transition from the R-attractor to the S-attractor, there is a phase in which $w_K>1$, i.e., the dominant energy condition $\rho_K>|p_K|$ is violated \cite{ArmendarizPicon:2000ah}. It means,  the k-essence energy can travel with superluminal speeds.  However, there are perfectly Lorentz-invariant theories with the non-standard kinetic term which allows the presence of superluminal speeds. Indeed, studies show that in spite of the superluminal propagation, the causal paradox does not arise in these theories and in this sense they are not less safe than General Relativity \cite{Babichev:2007dw,Babichev:2007zz}.

\section{Coupled dark energy and matter}\label{sec:coupledDE}

There is another attempt to solve the coincidence problem which is called coupled dark energy \cite{Wetterich:1994bg,Amendola:1999er,Das:2005yj,Zimdahl:2001ar,Chimento:2003iea,Cai:2004dk}. The dark energy density is of the same order as the dark matter energy density in the present Universe. Thus one could imagine that there is some connection between dark energy and dark matter.

The interaction between dark matter and dark energy in form of scalar field is described by following modified energy conservation equations
\begin{eqnarray}
\label{4.47}
\dot{\rho}_m+3H(\rho_m) & = & \delta,\\
\label{4.48}
\dot{\rho}_\phi+3H(\rho_\phi+p_\phi) & = & -\delta,
\end{eqnarray}
where $\delta$ is an energy exchange term in the dark sector.  The notation $m$ is for dark matter and $\phi$ is for dark energy.

Thus in brief,  constructing the coupled dark energy  model is finding an appropriate form of the coupling $\delta$. There are two major examples \cite{Amendola:1999er,Zimdahl:2001ar}:
\begin{eqnarray}
\delta & = & \kappa Q\rho_m\dot{\phi},\\
\delta & = & \alpha H(\rho_m+\rho_\phi),
\end{eqnarray}
where $Q$ and $\alpha$ are dimensionless constants. We use $\kappa^{2}=8\pi G$ in this section. Let us call those examples as coupling type 1 and coupling type 2.

\subsection{Coupling Type 1}

The first example is a coupled quintessence model assuming an exponential potential and linear coupling \cite{Amendola:1999er}.  The coupled quintessence scalar field equation is
\begin{equation}
\label{4.51}
\ddot{\phi}+3H\dot{\phi}+V_{,\phi}=-\kappa Q\rho_m,
\end{equation}
which is equivalent to $\dot{\rho}_\phi+3H(\rho_\phi+p_\phi)=-\kappa Q\rho_m\dot{\phi}$.
Here the potential is adopted by
\begin{equation}
V(\phi)=V_0 e^{-\kappa\lambda\phi}.
\end{equation}
The matter component behaves as
\begin{equation}
\dot{\rho}_m+3H(\rho_m)=\kappa Q\rho_m\dot{\phi},
\end{equation}
so that matter evolves as
\begin{equation}
\rho_m=\rho_{m0}a^{-3}e^{\kappa Q\phi}.
\end{equation}
The Friedmann equations for quintessence models are given by
\begin{equation}
\label{k}
H^{2}=\frac{\kappa^{2}}{3}\left(\frac{1}{2}\dot{\phi}^{2}+V+\rho_m+\rho_r\right)
\end{equation}

\begin{equation}
\label{k1}
\dot{H}=-\frac{\kappa^{2}}{2}\left(\dot{\phi}^{2}+\rho_m+\frac{4}{3}\rho_r\right).
\end{equation}
We will use the following  dimensionless variables \cite{Copeland:1997et,Copeland:2006wr,Tsujikawa:2010sc}
\begin{equation}
x=\frac{\kappa}{H}\frac{\dot{\phi}}{\sqrt{6}}, \indent y=\frac{\kappa}{H}\sqrt{\frac{V}{3}}, \indent z=\frac{\kappa}{H}\sqrt{\frac{\rho_r}{3}}.
\end{equation}
 $x^{2}$, $y^{2}$ and $z^{2}$ denote the energy density fraction carried by field kinetic energy, field potential energy and radiation, respectively. Thus, the density parameters are as follows  
\begin{equation}
\Omega_\phi=x^{2}+y^{2}, \indent \Omega_r=z^{2}, \indent \Omega_m=1-x^{2}-y^{2}-z^{2}.
\end{equation}
From the equations above (\ref{4.51}), (\ref{k}) and (\ref{k1}) one gets
\begin{eqnarray}
x^{\prime} & = & -3x+\frac{\sqrt{6}}{2}\lambda y^{2} +\frac{1}{2}x(3+3x^{2}-3y^{2}+z^{2})-\frac{\sqrt{6}}{2}Q(1-x^{2}-y^{2}-z^{2}),\\
y^{\prime} & = & -\frac{\sqrt{6}}{2}\lambda xy+\frac{1}{2}y(3+3x^{2}-3y^{2}+z^{2}),\\
z^{\prime} & = & -2z+\frac{1}{2}z(3+3x^{2}-3y^{2}+z^{2}),
\end{eqnarray}
where $^{\prime}\equiv \frac{d}{d \ln a}$ and $\lambda\equiv -\frac{V_{,\phi}}{\kappa V}$.

Now we can find the critical points (attractors) that satisfy $x^{\prime}=y^{\prime}=z^{\prime}=0$ on which the scalar field equation of state is
\begin{equation}
\label{ }
w_\phi=\frac{x^{2}-y^{2}}{x^{2}+y^{2}}=const..
\end{equation}
The effective equation of state $w_{eff}=p_{tot}/\rho_{tot}$ is
\begin{equation}
\label{ }
w_{eff}=x^{2}-y^{2}+\frac{z^{2}}{3}=\Omega_\phi w_\phi+\Omega_r w_r
\end{equation}
and $w_{eff}<-\frac{1}{3}$ is enough for cosmic acceleration.

There exist only two attractors which allow accelerated cosmic expansion. However, one of them does not contain the matter dominated era and thus fails to explain the large scale structure. So we consider just one attractor for this model. According to this scenario during the matter dominated era the scalar field has a finite and almost constant energy density. This field-matter-dominated era between the radiation era and the accelerated era is called $\phi MDE$ \cite{Amendola:1999er}. The $\phi MDE$ is characterized by
\begin{equation}
\label{ }
(x,y,z)=\left(-\frac{\sqrt{6}Q}{3},0,0\right), \indent \Omega_\phi=\frac{2Q^{2}}{3}, \indent w_\phi=1, \indent w_{eff}=\frac{2Q^{2}}{3}.
\end{equation}
The $\phi MDE$ is responsible for most of the differences with respect to the uncoupled quintessence model. For example, the evolution of the scale factor during the $\phi MDE$ is given by
\begin{equation}
\label{ }
a\propto t^{\frac{2}{3+2Q^{2}}}.
\end{equation}
Finally the field falls into the attractor which is characterized by
\begin{equation}
\label{ }
(x,y,z)=\left(\frac{\lambda}{\sqrt{6}},\sqrt{1-\frac{\lambda^{2}}{6}},0\right), \indent \Omega_\phi=1, \indent w_\phi=\frac{\lambda^{2}}{3}-1, \indent w_{eff}=\frac{\lambda^{2}}{3}-1.
\end{equation}
The scale factor evolves as
\begin{equation}
\label{ }
a\propto t^{\frac{2}{\lambda^{2}}}.
\end{equation}
This attractor causes accelerated expansion of the Universe for $\lambda^{2}<2$. Once this attractor is reached, matter density becomes zero. Hence according to this scenario the attractor is not yet reached at the present time, but the expansion is already accelerated.

The CMB data constrain the dimensionless coupling constant to be $|Q|<0.1$ \cite{Amendola:1999er}.

\subsection{Coupling Type 2}

In coupling type 1 they assumed  a specific potential and coupling from the outset. In coupling type 2, interaction potential and coupling structure are determined from the requirement $\frac{\rho_m}{\rho_\phi}=const.$ \cite{Zimdahl:2001ar}. 

The coupling equation (\ref{4.48}) is equivalent to
\begin{equation}
\label{4.68}
\dot{\phi}\left[\ddot{\phi}+3H\dot{\phi}+V_{,\phi}\right]=-\delta,
\end{equation}
where $_{,\phi}$ denotes derivative respect to $\phi$.
Defining the coupling as
\begin{equation}
\label{4.69}
\delta\equiv -3H\Pi_m\equiv 3H\Pi_\phi,
\end{equation} 
where  $\Pi_m=-\Pi_\phi$ is a relation between effective pressures, the equations (\ref{4.47}) and (\ref{4.48}) become
\begin{eqnarray}
\label{ }
\dot{\rho}_m+3H(\rho_m+\Pi_m) & = & 0,\\
\label{ }
\dot{\rho}_\phi+3H(\rho_\phi+p_\phi+\Pi_\phi) & = & 0.
\end{eqnarray}

Now we consider the requirement for an attractor solution. The time evolution of the ratio $\rho_m/\rho_\phi$ is
\begin{equation}
\label{ }
\left(\frac{\rho_m}{\rho_\phi}\right)^{\cdot}=\frac{\rho_m}{\rho_\phi}\left[\frac{\dot{\rho}_m}{\rho_m}-\frac{\dot{\rho}_\phi}{\rho_\phi}\right].
\end{equation}
Defining $\gamma_\phi\equiv\frac{\rho_\phi+p_\phi}{\rho_\phi}=\frac{\dot{\phi^{2}}}{\rho_\phi}$ and $\rho\equiv\rho_m+\rho_\phi$ we get
\begin{equation}
\label{4.73}
\left(\frac{\rho_m}{\rho_\phi}\right)^{\cdot}=-3H\frac{\rho_m}{\rho_\phi}\left[1-\gamma_\phi+\frac{\rho}{\rho_m\rho_\phi}\Pi_m\right].
\end{equation}
We can easily find a stationary solution 
\begin{equation}
\label{4.74}
\Pi_m=-\Pi_\phi=\frac{\rho_m\rho_\phi}{\rho}(\gamma_\phi-1)
\end{equation}
which satisfies $\left(\frac{\rho_m}{\rho_\phi}\right)^{\cdot}=0$.

Substituting the solution Eq.(\ref{4.74}) in the Eq.(\ref{4.69}) yields
\begin{equation}
\label{4.75}
\delta=-3H(\gamma_\phi-1)\frac{\rho_\phi\rho_m}{\rho}=-3H(\gamma_\phi-1)\frac{r}{r+1}\rho_\phi,
\end{equation}
where $r\equiv\frac{\rho_m}{\rho_\phi}=const.$.

Now we consider the stability of this stationary solution against small perturbation,
\begin{equation}
\label{ }
\frac{\rho_m}{\rho_\phi}=\left(\frac{\rho_m}{\rho_\phi}\right)_{st}+\epsilon.
\end{equation}
Then from the Eq.(\ref{4.73}) we get
\begin{equation}
\label{4.77}
\dot{\epsilon}=3H\left[\left(\frac{\rho_m}{\rho_\phi}\right)_{st}+\epsilon\right]\left[\frac{p_\phi}{\rho_\phi}-\frac{\rho}{\rho_\phi}\frac{\Pi_m}{\rho_m}\right]=3H\left[\left(\frac{\rho_m}{\rho_\phi}\right)_{st}+\epsilon\right]\left[\frac{p_\phi}{\rho_\phi}-\left(1+\left(\frac{\rho_m}{\rho_\phi}\right)_{st}+\epsilon\right)\frac{\Pi_m}{\rho_m}\right].
\end{equation}
We choose $\Pi_m=-c\rho$ where $c>0$ so that the interaction is symmetric in $\rho_m$ and $\rho_\phi$.  Up to the  first order in $\epsilon$ we have
\begin{equation}
\label{ }
\dot{\epsilon}=3Hc\frac{r^{2}-1}{r}\epsilon,
\end{equation}
so the stationary solution is stable for $r<1$. From the Eqs.(\ref{4.69}) and (\ref{4.75}), the value of the constant $c$ is
\begin{equation}
\label{ }
c=r\frac{1-\gamma_\phi}{(1+r)^{2}}
\end{equation}
and the constant $c$ has positive value for $\gamma_\phi<1$. Since $p\approx p_\phi$ today, the stability condition which follows from the first expression in (\ref{4.77}) corresponds to
\begin{equation}
\label{ }
\frac{p}{\rho}-\frac{\Pi_m}{\rho_m}\leq0.
\end{equation}
We seek a solution for negative pressure which implies $\gamma_\phi<1$. Combining this with the Eqs.(\ref{4.69}) and (\ref{4.75}), we have conditions : $\Pi_m<0$, $\delta<0$ which lead
\begin{equation}
\frac{|\Pi_m|}{\rho_m}\leq\frac{|p|}{\rho}.
\end{equation}

Connecting the Eqs.(\ref{4.47}), (\ref{4.48}) with the Eq.(\ref{4.75}), we get $\rho_m\propto\rho_\phi\propto a^{-3\frac{\gamma_\phi+r}{r+1}}$.  Putting it in the deceleration parameter (\ref{2.9}) implies that the power law accelerated expansion will occur for
\begin{equation}
\label{4.82}
r+3\gamma_\phi<2.
\end{equation}
Combining energy density evolution related to scale factor and the Friedmann equation, one gets the time evolution of the scale factor and the field energy density as well. Using the definition of $\gamma_\phi$, we find finally 
\begin{equation}
\label{ }
V(\phi)=\frac{1}{6\pi G}\left(1-\frac{\gamma_\phi}{2}\right)\frac{1+r}{(\gamma_\phi+r)^{2}}\frac{1}{t^{2}}
\end{equation}
and $V(\phi)_{,\phi}=-\lambda V(\phi)$, where $\lambda=\sqrt{\frac{24\pi G}{\gamma_\phi(1+r)}}(\gamma_\phi+r)$ so that
\begin{equation}
\label{ }
V(\phi)=V_0 e^{-\lambda(\phi-\phi_0)}.
\end{equation}
See Ref. \cite{Zimdahl:2001ar} for detailed calculation.

Defining an interaction potential $V_{int,\phi}\equiv\frac{\delta}{\dot{\phi}}$, one gets
\begin{equation}
\label{ }
V_{int}=-\frac{2r}{\gamma_\phi+r}\frac{1-\gamma_\phi}{2-\gamma_\phi}V(\phi).
\end{equation}
Introducing an effective potential
\begin{equation}
\label{ }
V_{eff}\equiv V(\phi)+V_{int},
\end{equation}
the coupling equation (\ref{4.68}) becomes
\begin{equation}
\label{ }
\ddot{\phi}+3H\dot{\phi}+V_{eff,\phi}=0.
\end{equation}
Considering the condition for accelerated expansion (\ref{4.82}), $\lambda$ is restricted to
\begin{equation}
\label{ }
\lambda^{2}<24\pi G\frac{(1-\gamma_\phi)^{2}}{(1+r)\gamma_\phi}.
\end{equation}

In this model the potential is not an input but derived from the coupling that satisfies required property of the attractor solution. But the explanation for how the interaction is exactly started  at the transition era from decelerated to accelerated expansion is still missing.

\subsection{Chameleon mechanism}
As mentioned in section~\ref{sec:quintessence},  the quintessence field mass must be of order of $H_0$. If we consider it as a coupled field with matter and assume the interaction is as strong as gravity, the coupling must be tuned to a small value ($|Q|<0.1$) to satisfy the test of the equivalence principle. The equivalence principle provides that, gravitational mass and inertial mass are the same and the laws of gravity are the same in any inertial frame. 

There is a suggestion  which allows scalar fields to have couplings to matter of order unity. In this scenario the  mass of the scalar field depends on the local matter density. In high density environment  like on the Earth the field is massive, but in the low density environment like in the solar system the field is essentially free. Such a scalar field is named chameleon \cite{Khoury:2003rn,Khoury:2003aq} as it changes properties to fit its surroundings.

The action of a chameleon scalar field $\phi$ is given by
\begin{equation}
\label{4.89}
\textsl{S}=\int d^4x\sqrt{-g}\left[-\frac{1}{16\pi G}R+\frac{1}{2}g^{\mu\nu}\partial_{\mu}\phi\partial_{\nu}\phi-V(\phi)\right]+\int d^4x \mathcal{L}_m(\psi_m^{(i)}, g_{\mu\nu}^{(i)}),
\end{equation}
where the first integration term is the same as the normal quintessence action and the second term is a matter term which is coupled to the chameleon scalar field. Each matter field $\psi_m^{(i)}$ couples to a metric $g_{\mu\nu}^{(i)}$ which is related to the Einstein frame metric $g_{\mu\nu}$ by a conformal transformation
\begin{equation}
\label{4.90 }
g_{\mu\nu}^{(i)}=e^{2\kappa\beta_i\phi}g_{\mu\nu},
\end{equation}
where $\beta_i$ are dimensionless constants. So the scalar field $\phi$ interacts with matter through a conformal coupling in form of $e^{\kappa\beta_i\phi}$. 

The field potential is assumed to be of the runaway form and satisfies $V\rightarrow 0$ as $\phi\rightarrow \infty$ and $|V_{,\phi}|\rightarrow \infty $ as $\phi\rightarrow 0$ such as
\begin{equation}
\label{4.91 }
V(\phi)=M^{4+n}\phi^{-n}
\end{equation}
which is familiar from the subsection~\ref{subsec:traker_sol}.

Varying the action (\ref{4.89}) with respect to $\phi$, a similar calculation to the Eq.(\ref{2.46}), we get the equation of motion for $\phi$
\begin{equation}
\label{4.92}
\partial_\nu\partial^{\nu}\phi=-V_{,\phi}-\sum_{i}\kappa\beta_i e^{4\kappa\beta_i\phi}g_{(i)}^{\mu\nu}T_{\mu\nu}^{(i)},
\end{equation}
where $T_{\mu\nu}^{(i)}=({2}/{\sqrt{-g^{(i)}}}){\delta\mathcal{L}_m}/{\delta g_{(i)}^{\mu\nu}}$ is the stress-energy tensor for the $i$th form of matter. For non-relativistic matter, we have $g_{(i)}^{\mu\nu}T_{\mu\nu}^{(i)}\approx\tilde{\rho}_i$, where $\tilde{\rho}_i$ is energy density. Introducing the energy density $\rho_i\equiv \tilde{\rho}_i e^{3\kappa\beta_i\phi}$ which is conserved in the Einstein frame, the Eq.(\ref{4.92}) reduces to
\begin{equation}
\label{4.93}
\partial_\nu\partial^{\nu}\phi=-V_{,\phi}-\sum_{i}\kappa\beta_i \rho_i e^{\kappa\beta_i\phi}.
\end{equation}
If we define the right hand side of this equation as $-V_{eff,\phi}$, the effective potential is dubbed as
\begin{equation}
\label{ }
V_{eff}(\phi)\equiv V(\phi)+\sum_{i} \rho_i e^{\kappa\beta_i\phi}.
\end{equation}
For monotonically decreasing $V(\phi)$ and positive $\beta_i$ the effective potential $V_{eff}$ has a minimum. See Figure 4.3.

\begin{figure}[htb]
  \centering
  \includegraphics[scale=0.7]{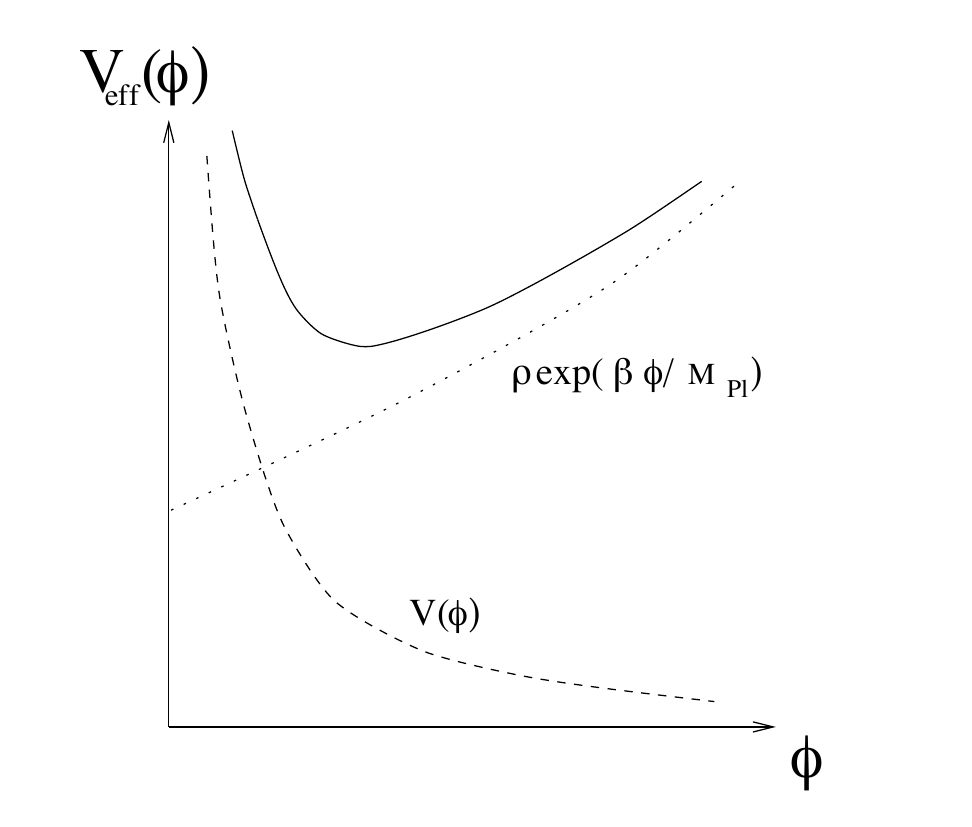}
  \caption[Kurzform f"ur das Abbildungsverzeichnis]{The chameleon effective potential $V_{eff}$ (solid curve) is the sum of two contributions. One from the actual potential $V_{\phi}$ (dashed curve), and the other from its coupling to the matter density $\rho$ (dotted curve). From Ref.(\cite{Khoury:2003rn}) }
\end{figure}

Calculating the point that satisfies $V_{eff,\phi}=0$ and $V_{eff,\phi\phi}=m^{2}$, we can define $\phi_{min}$ as the value of $\phi$ at the minimum and its mass $m_{min}$. Furthermore we recognize here that larger values of $\rho_i$ corresponds to smaller $\phi_{min}$ and larger $m_{min}$, i.e., the denser environment follows the more massive chameleon.

Let us consider a solution for a compact object which is static, spherically symmetric with radius $R_c$, homogeneous density $\rho_c$ and total mass $M_c=4\pi R_c^{3}\rho_c/3$. Then the Eq. (\ref{4.93}) becomes
\begin{equation}
\label{4.95}
\frac{d^{2}\phi}{dr^{2}}+\frac{2}{r}\frac{d\phi}{dr}=V_{,\phi}+\kappa\beta\rho(r)e^{\kappa\beta\phi},
\end{equation}
where all $\beta_i$ are assumed to be of the same value $\beta$. And $\rho(r)=\rho_c$ for $r<R_c$, $\rho(r)=\rho_\infty$ for $r>R_c$, where $\rho_\infty$ denotes the surrounding homogeneous matter density.

We define $\phi_c$ and $\phi_\infty$ as the value of $\phi$ that minimizes $V_{eff}$ with $\rho=\rho_c$ and $\rho_\infty$, respectively. And the corresponding masses of small fluctuations are $m_c$ and $m_\infty$. Boundary conditions are given by $d\phi/dr=0$ at $r=0$ so that the solution is non-singular at the origin and $\phi\rightarrow\phi_\infty$ as $r\rightarrow \infty$ so that the $\phi$-force on a test particle vanishes at infinity.

Now we consider a large object. Inside the object the field minimizes $V_{eff}$ at $\phi\approx\phi_c$. Outside the object the field evolves as $\phi\sim\exp(-m_\infty r)/r$ and approaches to $\phi_\infty$ for $r\gg R_c$. During the transition, at $r=R_c$, $\phi$ and $d\phi/dr$ should be continuous thus at this point  a thin shell of thickness $\Delta R_c$ below the surface is introduced where the field grows.  

Solving the Eq.(\ref{4.95}) for outside the object, one gets
\begin{equation}
\label{4.96}
\phi(r)\approx-\phi_\infty\left(1-\frac{\kappa(\phi_\infty-\phi_c)}{6\beta\Phi_c}\right)\frac{R_c e^{-m_\infty(r-R_c)}}{r}+\phi_\infty,
\end{equation}
where $\Phi_c=GM_c/R_c$ is the Newtonian potential of the object.

In case of the thin shell condition, $\Delta R_c/R_c\ll1$, and the inside of the object being much denser than the outside, $\phi_c\ll\phi_\infty$, the exterior solution (\ref{4.96}) becomes
\begin{equation}
\label{4.97}
\phi(r)\approx-\left(\frac{\kappa\beta}{4\pi}\right)\left(\frac{3\Delta R_c}{R_c}\right)\frac{M_c e^{-m_\infty(r-R_c)}}{r}+\phi_\infty,
\end{equation}
where $\Delta R_c/R_c$ can be adjusted as
\begin{equation}
\label{ }
\frac{\Delta R_c}{R_c}\approx \frac{\kappa(\phi_\infty-\phi_c)}{6\beta\Phi_c}.
\end{equation}

In case of small objects, in the sense of $\Delta R_c/R_c>1$, however, the thin shell condition is not satisfied.  Since their entire volume contributes to the $\phi$-field outside, the fraction of the infinitesimal volume in the thin shell to the total volume $3\Delta R_c/R_c$ becomes 1, so the exterior solution is
\begin{equation}
\label{4.99}
\phi(r)\approx-\left(\frac{\kappa\beta}{4\pi}\right)\frac{M_c e^{-m_\infty(r-R_c)}}{r}+\phi_\infty.
\end{equation}

Consider the potential $V(\phi)=M^{4+n}\phi^{-n}$, take $n=\beta=1$ and apply these to the Earth which must have a thin shell. It follows that the interaction range (which is anti-proportional to the field mass) is of order $1 mm$ on the Earth and of order $10-10^{4} AU$ in the solar system \cite{Khoury:2003rn}.

 The chameleon with thin shell effect satisfies tests of gravity in laboratory as well as from the solar system data \cite{Khoury:2003rn}. Moreover the chameleon mechanism predicts that the magnitude of equivalence principle violations and fifth force are much greater in space than on the Earth. 
 The chameleon force on a test particle of mass $M$ and coupling $\beta$ is given by
 \begin{equation}
\label{4.100}
\vec{F}_\phi=-\kappa\beta M\vec{\nabla}\phi,
\end{equation}
so the $\phi$ can be thought as a potential for the fifth force. 

On the Earth, calculating $\nabla\phi$ from the Eq. (\ref{4.97}) and substituting in the Eq.(\ref{4.100}) we get the fifth force on a test mass $M$ and coupling $\beta_i$ of magnitude
\begin{equation}
\label{ }
\vec{F}_\phi=-\frac{\kappa^{2}\beta\beta_i}{4\pi}\left(\frac{3\Delta R_\oplus}{R_\oplus}\right)\frac{M_{\oplus} M}{r^{2}},
\end{equation}
with the Earth radius $R_\oplus$ and the Earth mass $M_\oplus$. The Earth has a thin shell , $\Delta R_\oplus/R_\oplus\ll1$, hence the fifth force on the Earth is suppressed.

But in satellite in space, calculating the $\nabla\phi$ from the Eq.(\ref{4.99}) without thin shell and substituting in the Eq.(\ref{4.100}), we get the significant fifth force between two bodies of mass $M_1$ and $M_2$ and its coupling $\beta_1$ and $\beta_2$
\begin{equation}
\label{ }
\vec{F}_\phi=-\frac{\kappa^{2}\beta_1\beta_2}{4\pi}\frac{M_1M_2}{r^{2}}.
\end{equation}
Then the total  force, i.e., the gravitational force plus the chameleon-mediated fifth force between two masses  is given by \cite{Khoury:2003rn}
 \begin{equation}
\label{ }
|\vec{F}|=\frac{G M_1M_2}{r^{2}}(1+2\beta_1\beta_2),
\end{equation} 
 hence, the effective Newton's constant is $G_{eff}=G(1+2\beta_1\beta_2)$.
 
 In a similar way chameleon predicts an equivalence principle violation. We can read out the extra acceleration component $a_\phi=-\kappa\beta \vec{\nabla}\phi$ from the Eq.(\ref{4.100}). The E\"otv\"os parameter which denotes  relative difference in free-fall acceleration for two bodies of different composition is defined by
 \begin{equation}
\label{ }
\eta\equiv \frac{\Delta a}{a}=\frac{a_\phi}{a_N},
\end{equation}
where $a_N$ is the Newtonian acceleration on the Earth. If future gravity tests in satellite measure an effective Newton's constant which differs by order unity from the measured value on the Earth, or find an equivalence principle violating signal stronger than allowed by a laboratory experiment, as the chameleon cosmology predicts, it will be a strong candidate for dark energy.

\section{Unified dark energy and matter}\label{sec:unifiedDE}
\subsection{Chaplygin gas}
 To avoid the potential energy fine-tuning of the quintessence, instead of the form of potential, we take a change in the equation of state of the background fluid \cite{Kamenshchik:2001cp,Bento:2002ps}. This exotic background fluid, the so-called Chaplygin gas, has the following equation of state:
\begin{equation}
\label{4.105}
p=-\frac{A}{\rho^{\alpha}},
\end{equation}
where $A$ is a positive constant. Inserting this equation of state into the energy conservation equation (\ref{2.5}), we obtain
\begin{equation}
\label{4.106}
\rho(t)=\left[A+\frac{B}{a^{3(1+\alpha)}}\right]^{\frac{1}{1+\alpha}},
\end{equation}
where $B$ is an integration constant.

In the early epoch, $a\ll1$, the Chaplygin gas energy density behaves as $\rho\propto a^{-3}$ which corresponds to the matter dominated Universe. In the late epoch, $a\gg1$, the energy density behaves as $\rho\approx A^{1/(a+\alpha)}=const.$ which corresponds to the de Sitter Universe. Thus the single fluid, Chaplygin gas, behaves as dark matter in the early epoch and dark energy in the later epoch. That is why we call it a unified model of dark energy and dark matter, and it is thought to be an attractive feature  that   explains both dark sectors in terms of a single component.

The effective sound speed for the Chaplygin gas is given by
\begin{equation}
\label{4.107}
c_s^{2}=\frac{dp}{d\rho}=-\alpha w,
\end{equation}
 where $w$ is the equation of state parameter of the Chaplygin gas. Substituting above Eqs.(\ref{4.105}) and (\ref{4.106}) into the Eq.(\ref{4.107})we obtain
 \begin{equation}
\label{ }
c_s^{2}=\alpha \left[1+\frac{B/A}{a^{3(1+\alpha)}}\right]^{-1}.
\end{equation}
In the Chaplygin gas model the sound speed is small at the early epoch and becomes larger at the late epoch. The problem is, the large sound speed at the late epoch leads to growth of inhomogeneities. If so, we should have observed fluctuations or blow up in the matter power spectrum in the large-scale structures, but they are not observed. The observation constricts the upper bound on the values of $\alpha$ \cite{Sandvik:2002jz}
\begin{equation}
\label{ }
|\alpha|\leq10^{-5}.
\end{equation}
Thus, the Chaplygin gas model with $|\alpha|\gg10^{-5}$ is ruled out. Moreover the Chaplygin gas mode is indistinguishable from the $\Lambda$CDM model which implies $\alpha=-1$ and $w=const.=0$ in the early epoch and $ -1$ in the later epoch, and $c_s=-\alpha w=1$ at the present epoch.

\subsection{K-essence as unified dark energy}

The problem of the Chaplygin gas model can be avoided by constructing unified model of dark energy and dark matter  using the k-essence model \cite{Scherrer:2004au}.

For the Lagrangian density of k-essence, $p(X)$ as in the Eq.(\ref{4.19}), we assume that $p(X)$ can be expanded about its extremum at $X=X_0$ in the form
\begin{equation}
\label{4.110}
p(X)=p_0+p_2(X-X_0)^{2},
\end{equation}
where $p_0$ and $p_2$ are constants. The k-essence pressure $p_K=p$ and the energy density $\rho_K=2Xp_{,X}-p$ satisfy the energy conservation equation $\dot{\rho}_k+3H(\rho_K+p_K)=0$ and follow
\begin{equation}
\label{4.111}
(p_{,X}+2Xp_{,XX})\dot{X}+6Hp_{,X}X=0.
\end{equation}

Around the solution $X=X_0$ we introduce small perturbation $\epsilon\equiv\frac{(X-X_0)}{X_0}\ll1$. Substituting the Eq.(\ref{4.110}) into the Eq.(\ref{4.111}) and calculating up to linear order we obtain
\begin{equation}
\label{4.112}
\dot{\epsilon}=-3H\epsilon.
\end{equation}
From $\dot{\epsilon}=\dot{X}/X_0$ and the Eq.(\ref{4.112}) we obtain a solution
\begin{equation}
\label{4.113}
X=X_0[1+\epsilon_1(a/a_1)^{-3}]
\end{equation}
in terms of new constants $\epsilon_1$ and $a_1$. $X\approx X_0$ when $\epsilon_1(a/a_1)^{-3}\ll1$.

Plugging the Eq.(\ref{4.110}) into $\rho_K=2Xp_{,X}-p$ and using the solution (\ref{4.113}) around $X=X_0$, we get
\begin{equation}
\label{ }
p_K\simeq p_0, \indent \rho_K\simeq -p_0+4p_2X_0^{2}\epsilon_1(a/a_1)^{-3}.
\end{equation}
Now we obtain the equation of state for k-essence as unified dark matter
\begin{equation}
\label{ }
w_K\simeq-\left[1-\frac{4p_2}{p_0}X_0^{2}\epsilon_1\left(\frac{a}{a_1}\right)^{-3}\right]^{-1}.
\end{equation}
It follows $w_K\rightarrow0$ at the early epoch and $w_K\rightarrow-1$ at the late epoch.

The resulting effective speed of sound as defined in the Eq.(\ref{4.24}) is 
\begin{equation}
\label{ }
c_s^{2}\simeq\frac{1}{2}\epsilon_1\left(\frac{a}{a_1}\right)^{-3}\ll1,
\end{equation}
thus the large sound speed problem is avoided.

\section{f(R) gravity}\label{sec:f(R)}
For modified matter models we have considered a modification of the right hand side of the Einstein's equation $G_{\mu\nu}=8\pi GT_{\mu\nu}$. Now we consider  to modify the left hand side of the Einstein's equation. It means we do not search for matter causing cosmic acceleration anymore, but rather modify gravity itself. Therefore in the action we do not add a scalar field term as we have done in the Eq.(\ref{4.1}) and the Eq.(\ref{4.19}), but rather modify the Einstein-Hilbert action term.

To modify the Einstein-Hilbert action one may consider an action with higher-orders of curvature invariants, for example,
\begin{equation}
\label{ }
\textsl{S}=-\frac{1}{16\pi G}\int d^4x\sqrt{-g}(R+\alpha R^{2}+\beta R_{\mu\nu}R^{\mu\nu}+\gamma R^{3}+\cdots).
\end{equation}
The higher-order terms can be either of fundamental origin or they can arise as a result of vacuum polarization \cite{Mukhanov:2005sc}. 
In general Einstein gravity causes second order equations of motion. But every modification of the Einstein gravity introduces higher-derivative terms. I.e, in addition to the gravitational waves, the gravitational field has extra degrees of freedom including a scalar field \cite{Mukhanov:2005sc}. 

\subsection{f(R) cosmology}\label{sebsec:f(R)cosmology}
Consider an action
\begin{equation}
\label{5.2}
\textsl{S}=\frac{1}{16\pi G}\int d^4x\sqrt{-g}f(R)+\int d^4x \sqrt{-g} \mathcal{L}_M,
\end{equation}
where $f(R)$ is an arbitrary function of the scalar curvature $R$. Using a conformal transformation $g_{\mu\nu}\rightarrow \tilde{g}_{\mu\nu}=(\partial f/\partial R)g_{\mu\nu}$, it is possible to show that the higher derivative gravity theory ($Jordan$ $frame$) is conformally equivalent to Einstein gravity with an extra scalar field ($Einstein$ $frame$) \cite{Capozziello:2003tk,Mukhanov:2005sc}. We call this modified gravity theory as f(R) gravity \cite{Capozziello:2003tk,Carroll:2003wy,Nojiri:2006ri}. If the scalar field potential satisfies the slow-roll condition, then we have an inflationary solution. Thus f(R) gravity causes inflation due to the potential of a scalar field.

Varying the action (\ref{5.2}) with respect to the metric $g_{\mu\nu}$, it yields the modified Einstein equation
\begin{equation}
\label{5.3}
f_{,R}R_{\mu\nu}-\frac{1}{2}f g_{\mu\nu}-\nabla_\mu\nabla_\nu f_{,R}+g_{\mu\nu}\Box f_{,R}=8\pi GT_{\mu\nu},
\end{equation}
where $_{,R}$ denotes the derivative with respect to $R$, $\nabla_\mu$ is the covariant derivative and $\Box$ is the D'Alembert operator defined as $\Box=g^{\mu\nu}\nabla_\mu\nabla_\nu$. $T_{\mu\nu}$ is defined as in the Eq.(\ref{2.24}).  The trace of the Eq.(\ref{5.3}) is given by
\begin{equation}
\label{5.4}
3\Box f_{,R}+f_{,R}R-2f=8\pi GT,
\end{equation}
where $T=g^{\mu\nu}T_{\mu\nu}=\rho_M-3p_M$. 

To get an inflationary solution we consider the de Sitter point. The de Sitter space is a vacuum solution ($T=0$) with constant positive curvature, i.e., $R=const.$. Thus at this point $\Box f_{,R}=0$ so that
\begin{equation}
\label{5.5}
f_{,R}R-2f=0.
\end{equation}
If a dark energy model based on the f(R) gravity satisfies this condition, the late-time de Sitter solution can be realized.

Assuming a flat FLRW metric, we get modified Friedmann equations from the Eqs.(\ref{5.3}) and (\ref{5.4}) \cite{Tsujikawa:2007xu}:
\begin{eqnarray}
\label{5.6}
3f_{,R}H^{2} & = & 8\pi G\rho_m+(f_{,R}R-f)/2-3H\dot{f}_{,R},\\
\label{5.7}
2f_{,R}\dot{H} & = & -8\pi G\rho_m-\ddot{f}_{,R}+H\dot{f}_{,R},
\end{eqnarray}
where the dot means derivative with respect to the cosmic time $t$ and the radiation component is ignored. To compare f(R) gravity with observation we rewrite the Eqs. (\ref{5.6}) and (\ref{5.7}) as
\begin{eqnarray}
\label{5.8}
3AH^{2} & = & 8\pi G(\rho_m+\rho_{DE}),\\
\label{5.9}
-2A\dot{H} & = & 8\pi G(\rho_m+\rho_{DE}+p_{DE}),
\end{eqnarray}
where $A$ is some constant and $p_m=0$. $\rho_{DE}$ and $p_{DE}$ are defined by
\begin{eqnarray}
8\pi G\rho_{DE} & \equiv & (f_{,R}R-f)/2-3H\dot{f}_{,R}+3H^{2}(A-f_{,R}),\\
8\pi Gp_{DE} & \equiv &  \ddot{f}_{,R}+2H\dot{f}_{,R}-(f_{,R}R-f)/2-(3H^{2}+2\dot{H})(A-f_{,R})
\end{eqnarray}
so that the continuity equation (\ref{2.5}) holds for $\rho_{DE}$ and $p_{DE}$.
From Eqs. (\ref{5.8}) and (\ref{5.9}) we obtain
\begin{equation}
\label{ }
w_{DE}\equiv \frac{p_{DE}}{\rho_{DE}}=-\frac{2A\dot{H}+3AH^{2}}{3AH^{2}-8\pi G\rho_m}.
\end{equation}
In order to recover the standard matter era in the past we can choose $A=1$ in the Eq.(\ref{5.8}). The equation of state $w_{DE}$ can be smaller than -1, i.e., a phantom equation of state, before reaching the de Sitter attractor \cite{Tsujikawa:2007xu}.

\subsection{Cosmological and local gravity constraints}
The modification of gravity affects the large-scale structure as well as local gravity. Studying them, we can observationally distinguish between the f(R) gravity model and the $\Lambda$CDM model. At the same time we obtain constraints that the viable f(R) has to satisfy.

First of all, we need $f_{,R}(R)<0$ to avoid anti-gravity. And the effective scalar field mass of the f(R) gravity model is given by \cite{PhysRevD.75.044004}
\begin{equation}
\label{ }
M_{f(R)}^{2}\simeq\frac{1}{3f_{,RR}}
\end{equation}
in the regime $M_{f(R)}^{2}\gg |R|$.
Thus we need the condition $f_{,RR}(R)>0$ to avoid a tachyonic instability related with negative mass squared  \cite{PhysRevD.75.044004}. From those two stability conditions and the existence of a late-time  de Sitter point given by the Eq.(\ref{5.5}), it is stable for $0< \frac{Rf_{,RR}}{f_{,R}}\leq1$ \cite{Tsujikawa:2007xu}.

The first suggestion is $f(R)=-R+\alpha R^{2}$ \cite{Starobinsky:1980te}. But this model is not suitable to explain current accelerated expansion of the Universe because $R^{2}\ll |R|$ at the present epoch \cite{Tsujikawa:2007xu}. The next propose $f(R)=-R-\alpha/R^{n}$ $(\alpha>0, n>0)$ \cite{Carroll:2003wy} is also ruled out by the above constraints. 

The models that fulfill all such constraints are for example
\begin{eqnarray}
f(R) & = & -R-\mu R_c\frac{(R/R_c)^{2n}}{(R/R_c)^{2n}+1} \indent {\rm with} \indent  n,\mu, R_c>0,\\
f(R) & = & -R-\mu R_c\left[1-\left(1+\frac{R^{2}}{R_c^{2}}\right)^{-n}\right]  \indent{\rm with} \indent n,\mu, R_c>0,\\
f(R) & = & -R-\mu R_c\tanh \left(\frac{|R|}{R_c}\right)  \indent {\rm with} \indent  \mu, R_c>0.
\end{eqnarray}
These models satisfy the relation $f(R=0)=0$, i.e., the cosmological constant disappears in a flat space time.

The $\Lambda$CDM model and the f(R) gravity model predict different structure formation histories. For the large-scale structure observations we consider the wavenumber $k$ in sub-horizon scales $k/a\gg H$. By quasi-static approximation one gets the equation for matter density perturbation $\delta_m$  \cite{Tsujikawa:2007gd}
\begin{equation}
\label{ }
\ddot{\delta}_m+2H\dot{\delta}_m-4\pi G_{eff}\rho_m\delta_m\simeq0,
\end{equation}
where the effective gravitational constant $G_{eff}$ is defined by \cite{Tsujikawa:2007gd}
\begin{equation}
\label{ }
G_{eff}\equiv-\frac{G}{f_{,R}}\frac{1-4mk^{2}/(a^{2}R)}{1-3mk^{2}/(a^{2}R)}, \indent m\equiv \frac{Rf_{,RR}}{f_{,R}}.
\end{equation}
Here $m$ characterizes the deviation from the $\Lambda$CDM model which denotes $f(R)=-R-2\Lambda$.

If the deviation from $\Lambda$CDM is small, i.e., $-mk^{2}/(a^{2}R)\ll1$, then the effective gravitational coupling $G_{eff}$ is very close to the gravitational constant $G$ so that $\delta_m\propto t^{2/3}$ during the matter dominant epoch. But in the regime  $-mk^{2}/(a^{2}R)\gg1$, the effective gravitational coupling approaches $G_{eff}\simeq-4G/3f_{,R}$ so that $\delta_m\propto t^{(\sqrt{33}-1)/6}$ \cite{Tsujikawa:2007xu}. Later we will see that the DGP model implies the same aspect, i.e., on small scales we recover the Einstein gravity meanwhile on large scales the modified gravity becomes important. 

Computing the matter power spectrum $P_{\delta_m}=|\delta_m|^{2}$, we can justify whether the f(R) gravity realizes in nature. Moreover the modified evolution of matter perturbation directly affects the shear power spectrum in weak lensing \cite{Lombriser:2010mp}.
To be consistent with the local gravity constraints in the solar system, the function f(R) needs to be very close to that in the $\Lambda$CDM model in high density regions \cite{Hu:2007nk}. The SN Ia data constraint the deviation parameter $m$ to be $m(z=0) < 0.3$ \cite{Tsujikawa:2007xu}. In high density regions, $|R|\gg |R_0|$, the linear expansion of $R$ with respect to the cosmological value $R_0$ is no longer valid. In order to evade the solar system tests, one derived the non-linear chameleon mechanism \cite{Hu:2007nk}, where the chameleon mechanism may suppress the fifth forces in such a nonlinear regime.

\section{DGP model}\label{sec:DGP}
The brane world model of Dvali, Gabadadze and Porrati (DGP model) \cite{Dvali:2000hr} suggests that cosmic acceleration is a signal of  lack of understanding of gravitational interactions. Let us suppose that our 4-dimensional (4D) world is a $brane$ which is embedded in 5-dimensional (5D) $bulk$ Minkowski space-time with infinitely large extra dimensions. All particles and standard model forces are pinned onto the brane world, like dust particles on soap bubbles, while gravity is allowed to explore into the 5D bulk.
\subsection{Brane cosmology}
The action for the DGP model is given by
\begin{equation}
\label{ }
\textsl{S}=-\frac{M_{(5)}^{3}}{2}  \int d^5X\sqrt{-\tilde{g}}\tilde{R}-\frac{M_{pl}^{2}}{2}\int d^4x\sqrt{-g}R+\int d^4x \sqrt{-g} \mathcal{L}_m,
\end{equation}
where $\tilde{g}_{AB}$ is the metric in the bulk and $\tilde{R}$ is its Ricci scalar. And $g_{\mu\nu}$ is the induced metric on the brane and $R$ is the corresponding Ricci scalar. The first and second terms form Einstein-Hilbert actions in the 5D bulk and on the brane, respectively.  The third term is the matter action where $ \mathcal{L}_m$ is the matter Lagrangian confined to the brane. The second term may be generated by quantum corrections from the 5D gravity, or its coupling with a certain 5D scalar field \cite{Dvali:2000hr}.

Capital letters, superscripts and subscripts will be used for 5D quantities ($A,B=0,1,2,3,5$) whereas the 4D coordinates of the brane are $x_\mu$ ($\mu=0,1,2,3$). The extra coordinate will be denoted by $y$. The induced metric on the brane is given by
\begin{equation}
\label{ }
g_{\mu\nu}=\partial_\mu X^{A}\partial_\nu X^{B}\tilde{g}_{AB},
\end{equation}
\begin{equation}
\label{ }
g_{\mu\nu}(x)\equiv \tilde{g}_{\mu\nu}(x,y=0).
\end{equation}

We define the 4D Planck mass as $M_{pl}$ and the 5D Planck mass as $M_{(5)}$. The cross-over scale $r_c$ is defined by
\begin{equation}
\label{ }
r_c\equiv\frac{M_{pl}^{2}}{2M_{(5)}^{3}}.
\end{equation}
If the characteristic length scale $r\equiv\sqrt{x_1^{2}+x_2^{2}+x_3^{2}}$ is much smaller than the cross-over scale, $r_c$, gravity behaves as usual 4D theory. In contrast, at large distance gravity slips into the bulk so that it occurs a $weakening$ $of$ $gravity$ $on$ $the$ $brane$. Thus the higher dimension plays an important role. 

It can be easily understood through considering the brane as a metal sheet immersed in air \cite{Lue:2005ya}. Imagine that sound waves represent gravity. If one strikes the metal sheet the sound waves propagate along the metal sheet as well as into the air. However the energy of the sound wave on the sheet is much denser than in the air. Therefore an observer on the sheet does not feel the extra dimension (air) at all. Only after the wave has propagated a long distance so that an amount of sound wave energy has been lost into some unknown region, the observer becomes aware of the existence of a extra dimension.

Across the cross-over scale $r_c$, the weak-field gravitational potential behaves as \cite{Dvali:2000hr, Deffayet:2001pu} 
\begin{eqnarray}
\label{ }
V(r) \sim \left\{
\begin{array}{c}
r^{-1}\indent  {\rm for} \indent r\ll r_c,\\
r^{-2}\indent  {\rm for} \indent r\gg r_c.
 \end{array}\right.
\end{eqnarray}

How does the extra dimension change cosmology on the 4D brane? Let us consider the 5D space-time metric
\begin{equation}
ds^2=\tilde{g}_{AB}dx^{A}dx^{B}.
\end{equation}
Since we are interested in cosmology on the brane, the 5D line element is given by
\begin{equation}
	ds^2=N^2(t,y)dt^2-A^2(t,y)\gamma_{ij}dx^{i}dx^{j}-B^2(t,y)dy^2,
\end{equation}
where $\gamma_{ij}$ is a maximally symmetric 3D metric and the brane is a hyper-surface defined by $y=0$. The metric coefficients read \cite{Deffayet:2000uy}
\begin{eqnarray}
N(t,y) & = & 1+\epsilon|y|\ddot{a}(\dot{a}^2+k)^{-1/2},\\
A(t,y) & = & a+\epsilon|y|(\dot{a}^2+k)^{1/2},\\
B(t,y) & = & 1,
\end{eqnarray}
where $a(t)$ is the 4D scale factor and $\epsilon=\pm1$. Taking $y=0$, we obtain the usual 4D FLRW form.

The Einstein equation in 5D bulk is given by \cite{Deffayet:2000uy}
\begin{equation}
\tilde{G}_{AB}\equiv \tilde{R}_{AB}-\frac{1}{2}\tilde{R}\tilde{g}_{AB}=0,
\end{equation}
where $\tilde{G}_{AB}$ is 5D Einstein tensor. The induced 4D Einstein equation is as follows \cite{Deffayet:2000uy,Hinterbichler:2009kq}
\begin{equation}
G_{\mu\nu}-\frac{1}{r_c}(K_{\mu\nu}-Kg_{\mu\nu})=8\pi GT_{\mu\nu},
\end{equation}
where $K_{\mu\nu}$ is the extrinsic curvature on the brane. 
On the FLRW brane we get the modified Friedmann equation \cite{Deffayet:2000uy,Deffayet:2001pu}
\begin{equation}
\label{5.122}
H^2+\frac{k}{a^2}=\left(\sqrt{\frac{\rho}{3M_{pl}^2}+\frac{1}{4r_c^2}}+\frac{\epsilon}{2r_c}\right)^2,
\end{equation}
where $\rho$ denotes the total cosmic fluid energy density on the brane which satisfies the conservation equation (\ref{2.5}).
If $\frac{\rho}{3M_{pl}^2}\gg\frac{1}{4r_c^2}$, i.e., in the early Universe, this Friedmann equation takes the standard cosmology form: $H^2+\frac{k}{a^2}=\frac{8\pi G}{3}\rho$.

For a flat geometry ($k=0$) the Eq.( \ref{5.122}) takes the following form
\begin{equation}
\label{5.123}
H^2-\frac{\epsilon}{r_c}H=\frac{\rho}{3M_{pl}^2}.
\end{equation}
In case of $H^{-1}\ll r_c$ the second term of the Eq.(\ref{5.123}) is negligible so that we again recover the usual Friedmann equation: $H^2=\frac{8\pi G}{3}\rho$. On the other hand, in case of $H^{-1}\gg r_c$ the second term becomes important.

Depending on the sign of $\epsilon$, the cosmological solution Eq.(\ref{5.123}) has two different regimes. If $\epsilon=-1$ and $H^{-1}\gg r_c$, the Friedmann equation approaches
\begin{equation}
\label{ }
H\rightarrow H_\infty=\frac{\rho}{6M_{(5)}^{3}}.
\end{equation}
If $\epsilon=+1$, however, the matter dominated Universe ($\rho\propto a^{-3}$) approaches the de Sitter solution, 
\begin{equation}
H\rightarrow H_\infty=\frac{1}{r_c},
\end{equation}
which causes late-time acceleration. This cosmological solution is very interesting because it drives our Universe into $self$-$inflationary$ regime without dark energy.

Requiring that the late-time acceleration occurs around the present epoch, we get \cite{Deffayet:2001pu}
\begin{equation}
\label{ }
r_c\sim H_0^{-1},
\end{equation}
which corresponds to
\begin{equation}
\label{ }
M_{(5)}=10-100 MeV.
\end{equation}

\subsection{Observational test}

The most important task for observational tests of modified gravity is to distinguish it from the $\Lambda$CDM model. In order to discuss about cosmological tests it is convenient to rewrite the modified Friedmann equation (\ref{5.122}) in term of the redshift as \cite{Deffayet:2001pu}
\begin{equation}
\label{5.128}
H^{2}(z)=H_0^{2}\Bigg\{\Omega_k(1+z)^{2}+\left(\sqrt{\Omega_{r_c}}+\sqrt{\Omega_{r_c}+\Omega_m(1+z)^{3}}\right)^{2}\Bigg\},
\end{equation}
where the redshift is defined by $1+z= \frac{a_0}{a}$. And density parameters are defined by
\begin{equation}
\label{ }
\Omega_m \equiv \frac{8\pi G}{3}\frac{\rho_{m,0}}{ H_0^{2}a_0^{3}}, \indent \Omega_k\equiv \frac{-k}{H_0^{2}a_0^{2}}, \indent \Omega_{r_c}\equiv \frac{1}{4r_c^{2}H_0^{2}}.
\end{equation}
We compare this equation with the conventional Friedmann equation
\begin{equation}
\label{5.130}
H^{2}(z)=H_0^{2}\Big\{\Omega_k(1+z)^{2}+\Omega_m(1+z)^{3}+\Omega_{DE}(1+z)^{3(1+w_{DE})}\Big\}.
\end{equation}
Setting $z=0$ in the Eq.(\ref{5.128}), we obtain the normalization condition
\begin{equation}
\label{5.131}
\Omega_k+\left(\sqrt{\Omega_{r_c}}+\sqrt{\Omega_{r_c}+\Omega_m}\right)^{2}=1
\end{equation}
which differs from the conventional relation $\Omega_k+\Omega_m+\Omega_{DE}=1$.

For a flat Universe ($\Omega_k=0$) the Eq.(\ref{5.131}) becomes
\begin{equation}
\label{ }
\Omega_{r_c}=\left(\frac{1-\Omega_m}{2}\right)^{2}, \indent \Omega_{r_c}\leq1.
\end{equation}
Comparing the Eq.(\ref{5.128}) and the Eq.(\ref{5.130}), we see that $\Omega_{r_c}$ plays a role of $\Omega_{DE}$ but not same, as shown in the Eq.(\ref{5.131}). And for a flat Universe $\Omega_{r_c}$ is always smaller than  $\Omega_{DE}$.

Different Friedmann equations result in a different luminosity distance $d_L$  
\begin{equation}
\label{ }
d_L(z)=\frac{1+z}{\sqrt{ \Omega_k}H_0}S_k\left(H_0\sqrt{ \Omega_k}\int_0^{z}\frac{dx}{H(x)}\right),
\end{equation}
where
\begin{eqnarray}
\label{ }
S_k(x)= \left\{
\begin{array}{c}
\sin(x)\indent    ${\rm for}$ \indent k =1,\\
x\indent  \indent \   ${\rm  for}$ \indent k =0,\\
\sinh(x)\indent  ${\rm for}$ \indent k =-1.
 \end{array}\right.
\end{eqnarray}
The apparent magnitude in units of Mpc is given by
\begin{equation}
\label{ }
m=M+5 \log d_L+25,
\end{equation}
where $M$ is the absolute magnitude. Considering supernovae Ia as standard candles, $M$ is the same for all supernovae, and measuring apparent magnitudes of supernovae we can directly compare our model with SN Ia observations. To use this method we have to know the exact value of $H_0$.

In spite of attractive features,  the DGP model suffers from observational disfavor and ghost instabilities. SN Ia, BAO and CMB data shows that the modified Friedmann equation is less consistent with observations than the usual Friedmann equation, i.e., $\Lambda$CDM model \cite{Xia:2009gb}.  There is a modified version of the DGP model which is characterized by \cite{Dvali:2003rk}
\begin{equation}
\label{ }
H^{2}-\frac{H^{\alpha}}{r_c^{2-\alpha}}=\frac{8\pi G}{3}\rho
\end{equation}
so that the effective equation of state $w_{eff}$ evolves from $-1+\frac{2\alpha}{3}$ during the radiation dominated era to $-1+\frac{\alpha}{2}$, during the matter dominated era to $-1$ in the distant future. Recent observational data have constrained $\alpha=0.254\pm0.153$ \cite{Xia:2009gb}, thus the flat DGP model ($\alpha=1$) is ruled out.

According to the studies of the linear theory about a flat multidimensional space-time and a flat brane, the DGP model has a scalar ghost field localized near the brane \cite{Dubovsky:2002jm}. However it is possible to get a  ghost-free DGP model by embedding our visible 3D brane within a 4D brane in a flat 6D bulk \cite{deRham:2008zz}. Moreover Galilean gravity also gives rise to the possibility of avoiding the ghost problem \cite{Nicolis:2008in}.

\section{Inhomogeneous LTB model}\label{sec:LTB}
The apparent accelerated expansion of the Universe may not caused by dark energy but rather by inhomogeneities in the distribution of matter. The basic idea is that, there are inhomogeneities on a larger scale, in the form of underdense bubbles, i.e., we live in an underdense region of the Universe and describe its behavior, a faster expansion compared to the outside, as an apparent cosmic acceleration. 

How do we understand the concept of $apparent$ $cosmic$ $acceleration$? In a homogeneous Universe the expansion rate is a function of time only. But in an inhomogeneous Universe the expansion rate varies with both time and space. Thus, if one observes faster expansion rate for low redshifts than higher redshifts, in the homogeneous case it is described as cosmic acceleration, whereas in the inhomogeneous case it is a result of spatial variation with an expansion rate being faster as being closer to us \cite{Alnes:2005rw}.

The first underdense void model is proposed by Tomita \cite{Tomita:2000jj} in form of a local homogeneous void separated from the homogeneous FLRW outside with a singular mass shell. There occurs a discontinuous jump at the location of the mass shell. This model is extended  to a more realistic model with continuous transition between inside and outside of the void \cite{Alnes:2005rw}.  In this extended model the Universe is dust dominant and isotropic but inhomogeneous. The inhomogeneity is spherically symmetric. This model can be described by the Lema\^{i}tre-Tolman-Bondi (LTB) spherically symmetric models \cite{Lemaitre:1933gd,Tolman:1934za,Bondi:1947av}.

We have used the FLRW metric under the assumption that our Universe is isotropic and homogeneous. Now we renounce the homogeneity, then the line element for a spherically symmetric inhomogeneous Universe is given by
\begin{equation}
\label{6.1}
ds^{2}=dt^{2}-X^{2}(r,t)dr^{2}-R^{2}(r,t)d\Omega^{2}.
\end{equation}
For the Einstein equation we assume $T_{\mu\nu}=$ diag$(\rho,0,0,0)$, i.e., containing matter only.
Solving the equation $G_{01}=0$ gives
\begin{equation}
\label{ }
X(r,t)=\frac{R^{\prime}(r,t)}{\sqrt{1+\beta(r)}},
\end{equation}
where prime denotes derivative with respect to $r$ and $\beta(r)$ is a function of $r$. The metric (\ref{6.1}) recovers the FLRW metric by choosing $R=a(t)r$ and $\beta=-kr^{2}$.

The Einstein equations for the dust dominated LTB Universe are given by \cite{Alnes:2005rw}
\begin{eqnarray}
\label{6.3}
H^{2}_\bot+2H_\|H_\bot-\frac{\beta}{R^{2}}-\frac{\beta^{\prime}}{RR^{\prime}} & = & 8\pi G\rho, \\
\label{6.4}
6\frac{\ddot{R}}{R}+2H^{2}_\bot-2\frac{\beta}{R^{2}}-2H_\|H_\bot+\frac{\beta^{\prime}}{RR^{\prime}} & = & -8\pi G\rho,
\end{eqnarray}
where dot denotes derivative with respect to $t$. The transverse Hubble function is defined by $H_\bot\equiv \dot{R}/R$ and the radial Hubble function is defined by $H_\|\equiv \dot{R}^{\prime}/R^{\prime}$. Adding Eqs. (\ref{6.3}) and (\ref{6.4}), we obtain 
\begin{equation}
\label{6.5}
2R\ddot{R}+\dot{R}^{2}=\beta,
\end{equation}
and integrating this equation leads to
\begin{equation}
\label{6.6}
H^{2}_\bot=\frac{\alpha}{R^{3}}+\frac{\beta}{R^{2}},
\end{equation}
where $\alpha$ is a function of $r$. Thus the dynamical effect of $\alpha$ and $\beta$ are similar to those of dust and curvature, respectively.

We define the deceleration parameter $q_\bot\equiv-R\ddot{R}/\dot{R}^{2}$. See the Eq.(\ref{2.9}). Substituting Eqs.(\ref{6.5}) and (\ref{6.6}) into this expression yields
\begin{equation}
\label{ }
q_\bot=\frac{1}{2}\frac{\alpha}{\alpha+\beta R}.
\end{equation}
Since $\alpha\geq0$, the deceleration parameter has a positive value. Indeed for a spatially flat and dust dominated Universe, $q_\bot=0.5$. Thus a dust dominated  inhomogeneous Universe can not be accelerating.

We define $t=0$ as the time when photons decoupled from matter ($z\simeq1090$) and $R(r,t=0)=0$. And we introduce a conformal time $\eta$ by $d\eta=(\sqrt{\beta}/R)dt$. Integrating Eqs. (\ref{6.3}) and (\ref{6.4}) with $\beta>0$, we get a solution of the Einstein equation for the LTB model \cite{Alnes:2005rw}:
\begin{eqnarray}
R & = & \frac{\alpha}{2\beta}(\cosh\eta-1)  =  \frac{\Omega_{m,0}r}{2\Omega_{k,o}}(\cosh\eta-1), \\
t & =&\frac{\alpha}{2\beta^{3/2}}(\sinh\eta-\eta)  =  \frac{\Omega_{m,0}}{2H_{\bot,0}\Omega^{3/2}_{k,o}}(\sinh\eta-\eta),
\end{eqnarray}
where we chose $\alpha=H_{\bot,0}^{2}\Omega_{m,0}r^{3}$ and $\beta=H_{\bot,0}^{2}\Omega_{k,0}r^{2}$.

The structure of the underdence void causing apparent acceleration can be expressed as  \cite{GarciaBellido:2008nz}
\begin{eqnarray}
\Omega_{m,0}(r) & = & \Omega_{out}+(\Omega_{in}-\Omega_{out})\left(\frac{1-\tanh[(r-r_0)/2\Delta r]}{1+\tanh[r_0/2\Delta r]}\right), \\
H_{\bot,0}(r) & = & H_{out}+(H_{in}-H_{out})\left(\frac{1-\tanh[(r-r_0)/2\Delta r]}{1+\tanh[r_0/2\Delta r]}\right),
\end{eqnarray}
where $in$ and $out$ represent quantities inside and outside the void, respectively. Further $r_0$ characterizes the size of the void  and $\Delta r$ is the thickness of the transition shell.

From the CMB acoustic peak the value of the local density parameter $\Omega_{in}$ is constraint to be in the range 0.1-0.3, whereas $\Omega_{out}=1$ for asymptotic flatness. The HST observations imply $H_{in}\approx70$ km/sec/Mpc, whereas outside the void one requires $H_{out}\approx50$ km/sec/Mpc to be consistent with $\Omega_{out}$.  The SN Ia data constraint $r_0$ and $\Delta r$ to be $r_0=2.3\pm0.9$ Gpc and $\Delta r/r_0>0.2$, respectively  \cite{GarciaBellido:2008nz}.

We define relative matter and curvature densities from the Eq.(\ref{6.3}) as
\begin{eqnarray}
\Omega_m & = & \frac{8\pi G\rho}{H^{2}_\bot+2H_\|H_\bot}, \\
\Omega_k & = & 1-\Omega_m.
\end{eqnarray}
The usual definition (\ref{2.13}) for the homogeneous case can be recoverd with $H_\bot=H_\|$.

To compare the inhomogeneous LTB model with SN Ia observations, we need to find the luminosity distance. Photons which travel along radial null geodesics, $ds^{2}=d\Omega^{2}=0$, arriving at $r=0$ today $t_0$ follow a path $t=\hat{t}(r)$ given by
\begin{equation}
\label{ }
\frac{d\hat{t}}{dr}=-\frac{R^{\prime}(r,\hat{t})}{\sqrt{1+\beta}},
\end{equation}
with $\hat{t}(0)=t_0$. The redshift $z=z(r)$ of photons obeys the differential equation \cite{Iguchi:2001sq}
\begin{equation}
\label{ }
\frac{dz}{dr}=(1+z)\frac{\dot{R}^{\prime}(r,\hat{t})}{\sqrt{1+\beta}},
\end{equation}
with $z(0)=0$. Now we have the functions $\hat{t}(z)$ and $r(z)$. From that we can obtain the luminosity distance $d_L$ given by 
\begin{equation}
\label{ }
d_L(z)=(1+z)^{2}R(r,\hat{t}),
\end{equation}
which is related to the angular diameter distance $d_A(z)=R(r,\hat{t})$.

The inhomogeneous LTB model matches to the supernovae data and the location of the first acoustic peak of CMB temperature power spectrum \cite{Alnes:2005rw} but it is still challenging to reproduce the entire CMB angular power spectrum. The observed isotropy of the CMB radiation implies that we must live close to the center of the inhomogeneity. The maximum distance to the center is constrained to be smaller than 15 Mpc \cite{Alnes:2006pf}. I.e., we are located at special point in the Universe. This reminds us again of coincidence and the anthropic like question, why are we special?

\section{Summary}\label{sec:summary}

We summarize and compare the various dark energy models with respect to attractive features, rising problems and possible solutions.

  \subsubsection* {Cosmological constant}
  
  The simplest explanation for dark energy is the energy associated with the vacuum which has constant equation of state parameter $w=-1$.
  The cosmological constant has perfect fit to observations as shown in Figure 1.1. But it suffers two cosmological problems.
  The observed dark energy density value is 120 orders of magnitude smaller than the theoretical expectation (fine-tuning problem). Moreover dark energy density and dark matter energy density are of the same order today (coincidence problem).
  
  \subsubsection* {Quintessence}
  
  Dark energy could exist in the form of a scalar field which has time varying equation of state. According to the tracker solution, the quintessence component tracks the equation of state of the background (radiation in the radiation dominant epoch and matter in the matter dominant epoch) and only recently grows to dominate the energy density. The tracker behavior allows the quintessence model to be insensitive to initial conditions.
  But it needs fine-tuning of the potential energy as $\sqrt{V''(\phi)}\sim H_0\sim 10^{-33}eV$. Moreover, the quintessence model does not solve the first cosmological constant problem.
  
  \subsubsection* {K-essence}
  
  In the k-essence model, the non-canonical kinetic energy term in the Lagrangian can drive late-time cosmic acceleration without the help of potential energy.
  The attractor behavior allows the k-essence model to be insensitive to initial conditions as well. Unlike the quintessence model, the k-essence field tracks only the radiation background so that it is free from fine-tuning which arose in the quintessence model.
  Moreover the coincidence problem is solved by presence of an S-attractor which attracts shortly after the onset of the matter dominated phase. However the k-essence model does not explain the smallness of the vacuum energy.
  There is a superluminal sound speed phase, but a causal paradox does not arise. 
  
 \subsubsection* {Coupled dark energy}
  Dark energy density and dark matter energy density are of the same order today (coincidence problem), thus it is natural to imagine that dark energy and dark matter have some relation. Coupled dark energy models connect dark matter and dark energy. Hence dark sectors exchange energy with each other. But observations constraint the coupling to be small (coupling constant $|Q|<0.1$). This constraint may be avoided by adopting the chameleon mechanism in which a scalar field has couplings to matter of order unity and changes its mass to fit to the local matter density. It is expected that the measurement of effective gravitational constant in satellite project test this model.

 \subsubsection* {Unified models of dark energy and dark matter}
  
  These models explain both dark sectors in terms of a single component by using a Chaplygin gas with exotic equation of state. The Chaplygin gas behaves as dark matter in the early epoch and as dark energy in the later epoch.
 But it has large sound speed at late epoch which leads to growth of inhomogeneities so that it is ruled out by observation.
  The large sound speed problem can be avoided by introducing k-essence as unified dark matter which provides also $w\rightarrow 0$ at the early epoch and $w\rightarrow -1$ at the late epoch. 
  
  \subsubsection* {f(R) gravity}
  
  f(R) gravity modifies Einstein gravity. The higher order gravity terms (in form of function f(R)) in the Lagrangian are equivalent to Einstein gravity with an extra scalar field. If this scalar field satisfies some condition it can drive cosmic acceleration as well. So the f(R) gravity is thought to be equivalent to the quintessence.
  There are several models of  function f(R) which fulfill stability conditions as discussed in section~\ref{sec:f(R)}. Observing the matter power spectrum or weak lensing, we can justify whether f(R) gravity realizes in nature.
 The f(R) gravity model is strongly constraint by solar system tests. Adopting the non-linear chameleon mechanism, it is possible to build a viable model.
 
   \subsubsection* {DGP model}
  
  According to the DGP model, we are living on 4D brane world in a 5D extra dimensional bulk. All particles and standard model forces are pinned onto the brane world, only gravity can explore into the 5D bulk. It appears in large scale as weakening of gravity on the brane.
  The modified Friedmann equation in the DGP model allows self-acceleration of the Universe.
  It may explain the weakness of gravity compared to other forces and gives a connection to string theory. Moreover the DGP model can solve the first cosmological constant problem, because gravity in 5D bulk is screened into 4D brane.
  But this model is disfavored by observations and suffers from a ghost problem.
  The ghost problem may be avoided by introducing a 6D theory or Galilean gravity.
  
  \subsubsection* {Inhomogeneous LTB model}
  The inhomogeneous LTB model implies that current cosmic acceleration is not real at all. In other models, we have assumed FLRW metric, i.e. a homogeneous isotropic Universe. In  the inhomogeneous LTB model we drop out the homogeneity, but there is a huge underdense void. If one measures a larger expansion rate (weaker gravity in underdence region) for lower redshift than higher redshift, it appears as cosmic acceleration.
We may live in an underdense void which leads to this apparent cosmic acceleration.  
In large-scale structures we indeed observe inhomogeneous voids.
But it is still challenging to satisfy CMB constraints.
And this model implies that we are located at a special point (within 15 Mpc from the center of void).
\vspace{.5cm}

The current status for probing dark energy is as follows. All current data provide strong evidence for an accelerated expansion of the Universe. SNe, CMB and BAO data provide independent evidence that the Universe contains 73\% dark energy which only came to be dominant after the observed structure had formed. How can we explain this fact? It is an important issue in physics and cosmology today, what is the nature of dark energy and its time evolution if it exists. Is dark energy actually a fine-tuned vacuum energy and are we just lucky (as anthropic consideration)? Do quintessence, k-essence, coupled or unified models, modified gravity models or inhomogeneity models give better explanations? The k-essence model is one of the most self consistent models. The DGP model or the inhomogeneous LTB model may bring us a completely different view of the world in future. We do not have one promising model yet. Will the Universe recollapse or continue to expand with or without acceleration? The destiny of the Universe depends definitely on the nature of dark energy. 

There are many space- and ground-based observations in progress or being planned. They are aiming at selecting an appropriate dark energy model. It is still open, how we reduce systematic errors for observations.  To get more reliable observational data we need huge volumes probed by the most recent deep redshift surveys as well. We have characterized dark energy and its effects on the expansion through the equation of state parameter $w$ which is not what we actually measure. Thus it is convenient to introduce improved descriptions of dark energy which are better matched to observation. And we need theoretical models that make specific enough predictions to select out the better matching model or distinguish them from the $\Lambda$CDM model. For example, the chameleon mechanism provides very specific  predictions for the effective Newtonian constant and  the E\"otv\"os parameter in satellite measurement, as we discussed in section~\ref{sec:coupledDE}.

From Einstein's biggest blunder to the Nobel Prize in Physics 2011: now we are ready to meet the most mysterious cosmological discoveries.

\section*{Acknowledgments}
The authors wish to thank Viatcheslav Mukhanov for suggesting to review the subject. Y.W. would like to thank Nico Wintergerst for important comments on the earlier version of the draft. This work is supported by the TRR 33 ``The Dark Universe".

\appendix
\section{Basics of Cosmology}\label{sec:basics}
\subsection{Homogeneous FLRW model}\label{subsec:FLRW}

We will introduce here some cosmological quantities.
The cosmological dynamics are given by solving the Einstein equation
\begin{equation}
G_{\mu\nu}=8\pi GT_{\mu\nu},
\end{equation}
where the Einstein tensor $G_{\mu\nu}$ is defined as
\begin{equation}
G_{\mu\nu}\equiv R_{\mu\nu}-\frac{1}{2}g_{\mu\nu}R.
\end{equation}

Assuming the Universe is spatially homogeneous and isotropic its metric takes the Friedmann-Lema\^itre-Robertson-Walker (FLRW) form
\begin{equation}
ds^2\equiv g_{\mu\nu}dx^{\mu}dx^{\nu}=dt^2-a^2(t)\left[\frac{dr^2}{1-k r^2}+r^2d\Omega^2\right],
\end{equation}
where $d\Omega^2=d\theta^2 + \sin^2\theta d\phi^2$ and the sign function of the curvature parameter $k$ takes values +1, 0 and -1 for positively curved, flat and negatively curved spatial section, respectively.

We use the scale factor which is the relative size of the spatial sections as a function of time $a(t)=R(t)/R_0$, where the subscript $0$ represents the present time.
The scale factor can be expressed in terms of the redshift $z$ as $a=\frac{1}{1+z}$ normalized so that $a_0=1$.

On large scales matter in the Universe may be modeled as a perfect fluid which is characterized by energy density $\rho$, isotropic pressure $p$ and four-velocity $u_\mu\equiv\frac{dx_{\mu}}{ds}$ normalized so that $u^{\mu}u_{\mu}=1$. The energy-momentum tensor is given by
\begin{equation}
T_{\mu\nu}=(\rho+p)u_\mu u_\nu-pg_{\mu\nu}.
\end{equation}
The energy conservation equation ($T^{\alpha}_{\ 0;\alpha}=0$, where $_{;\alpha}$ denotes the covariant derivative with respect to $\alpha$ component) in an isotropic and homogeneous universe implies
\begin{equation}
\label{2.5}
\dot{\rho}=-3H(\rho+p),
\end{equation}
where the Hubble parameter is defined as
\begin{equation}
H\equiv \frac{\dot{a}}{a}.
\end{equation}
Another convenient dimensionless parameter that characterizes the expansion is the deceleration parameter:
 \begin{equation}
\label{2.9}
q\equiv -\frac{\ddot{a}}{aH^{2}}.
\end{equation}
For the accelerating Universe $q$ is defined to be negative.

Substituting the metric and the energy momentum tensor into the Einstein equation, we get the first and second Friedmann equations:
\begin{equation}
\label{2.6}
H^2+\frac{k}{a^2}=\frac{8\pi G}{3}\rho
\end{equation}
and 
\begin{equation}
\label{2.7}
\frac{\ddot{a}}{a}=-\frac{4\pi G}{3}(\rho+3p).
\end{equation}

For the relation between energy density $\rho$ and pressure $p$, we define the equation of state as
\begin{equation}
\label{ }
w\equiv \frac{p}{\rho}.
\end{equation}
 For example, non-relativistic cold matter (e.g. dust) is pressureless and corresponds to $w=0$, radiation to $w =\frac{1}{3}$, and the cosmological constant to $w=-1$. Combining the equation of state and the second Friedmann equation, we get the condition $-1\leq w<-\frac{1}{3}$ for the observational status, namely the accelerated expansion of the Universe. In general relativity, the $null$ $dominant$ $energy$ $condition$ ensures that energy does not propagate outside the light cone, i.e., $|p|\leq|\rho|$ which leads to $-1\leq w\leq1$. However, we see  that some models violate such conditions, phantom ($w<-1$) or superluminal sound speed ($w>1$).

For a flat universe ($k=0$), solving the Eqs.(\ref{2.5}) and (\ref{2.6}) or (\ref{2.7}), we express cosmological quantities in terms of $w$ as
\begin{equation}
\label{ }
\rho\propto a^{-3(1+w)},\qquad
a(t)\propto t^{\frac{2}{3(1+w)}},\qquad
H=\frac{2}{3(1+w)t},
\end{equation}
where $w$ is assumed to be time independent and $w \neq -1$. For $w=-1$,
\begin{equation}
\rho= {\rm const.},\qquad
a(t)\propto e^{Ht},\qquad
H={\rm const.}
\end{equation}

The total energy density can be divided into four components (cold matter, radiation, dark energy and curvature density)
\begin{equation}\label{eq:rho_tot}
\rho_{tot}= \rho_m +\rho_{r}+\rho_{\Lambda}+\rho_{k}
\end{equation}
and each component decays as powers of the scale factor $a$:
\begin{equation}\label{eq:rho_tot_0}
\rho_{tot}= \rho_{m,0}\left(\frac{a_0}{a}\right)^{3} +\rho_{r,0}\left(\frac{a_0}{a}\right)^{4}+\rho_{\Lambda,0}+\rho_{k,0}\left(\frac{a_0}{a}\right)^{2},
\end{equation}
where $\rho_k \equiv - 3k/(8\pi G a^2)$ and $\rho_{i,0}\equiv \rho_i(t_0)$. Here we have assumed $w=-1$ for dark energy component.
One can see that $\rho_{\Lambda}$ becomes relatively important as time goes by.

Now we define the present density parameters as
\begin{equation}
\label{2.13}
\Omega_i\equiv \frac{\rho_{i}(t_0)}{\rho_{cr}(t_0)},
\end{equation}
 where the critical density $\rho_{cr}\equiv\frac{3H^{2}}{8\pi G}$ is the total energy density of matter or energy needed for the Universe to be spatially flat. It follows from Eqs.~(\ref{2.6}) and (\ref{eq:rho_tot_0}) that
 \begin{equation}
\label{ }
1=\sum_{i}\Omega_i,
\end{equation}
where $\Omega_k \equiv -k/(a_0^2H_0^2)$ and $\Omega_k \simeq 0$ from figure~\ref{observation}.

In the Eq.~(\ref{eq:rho_tot}) we have introduced the $\Lambda$CDM model, which is impressively consistent with current observations.
The $\Lambda$CDM model is a cold dark matter model with dark energy, which has been proposed to explain observational results such as anisotropies of the CMB, the large scale structure of galaxy clusters, accelerating expansion and chemical distribution of the Universe.

The letter $\Lambda$ stands for the dark energy component which occupies $72.8^{+1.5}_{-1.6}$\% of the total energy density of the Universe. The cold dark matter is non-relativistic and  non-baryonic, and it does not interact with itself or other particles but interacts only through gravity. It is thought as making up $22.7\pm1.4$\% of our Universe. The remaining $4.56\pm0.16$\% is baryonic matter which builds visible planets, stars and galaxies. The current Hubble parameter is $H_0=70.4^{+1.3}_{-1.4}$ km/s/Mpc (From WMAP+BAO+$H_0$ data which are released at 2011 \cite{Jarosik:2010iu}).

The $\Lambda$CDM model is built up on the homogeneous, isotropic and flat Universe. In the $\Lambda$CDM model, the Universe was hot and dense and had an early phase of expansion, when the light elements are produced via big bang nucleosynthesis and the radiation plasma dominates the total energy density of the Universe. Before the radiation dominated phase, the Universe had an earlier epoch of accelerated expansion, called {\it inflation}, when primordial density perturbations were seeded by quantum fluctuations, leaving an imprint on the CMB anisotropy and leading to the formation of structures due to gravitational instability on the present, isotropic and flat Universe.

\subsection{Scalar field theory}

 There are several dark energy models which are based on a scalar field such as quintessence and k-essence. In this subsection we introduce the basics of scalar field theory which we need for understanding such models.

A scalar field is a field which is invariant under any Lorentz transformation and has a scalar value at any point in space. For example, the temperature of a swimming pool is a scalar field, i.e., at each point we can measure a scalar value of temperature. The scalar field is known as spin-zero particles, and recently the Higgs particle which is a scalar field has been observed at LHC. Scalar fields are often introduced because of mathematical simplicity.
To combine classical fields and general relativity we use the $principle$ $of$ $the$ $least$ $action$. The action $\textsl{S}$ is the time integral of the Lagrangian $L$ 
\begin{equation}
\textsl{S}\left[x_i\right]=\int L\left[x_i(t),\dot{x}_i(t)\right] dt
\end{equation}
and can be rewritten by integrating the Lagrangian density $\mathcal{L}$ over all space-time
\begin{equation}
\textsl{S}\left[\varphi_i\right]=\int \mathcal L\left[ \varphi_i(x)\right]d^4x,
\end{equation}
where $\varphi_i(x)$ are classical fields including scalar and metric fields. In classical mechanics the Lagrangian is defined as the kinetic energy of the system minus its potential energy
\begin{equation}
L=T-V.
\end{equation}
For classical fields, the Lagrangian is very useful because it is invariant under some transformations that represent symmetries of the system. We can directly obtain the Euler-Lagrange equation by taking the least action :
\begin{equation}
\frac{\delta \textsl{S}}{\delta x_i}=\frac{\partial L}{\partial {x}_i}-\frac{d}{dt}\left(\frac{\partial L}{\partial \dot{x}_i}\right)=0.
\end{equation}

The Einstein-Hilbert action describing gravity in general relativity is given by
\begin{equation}
\textsl{S}=-\frac{1}{16\pi G}\int R\sqrt{-g}d^4x,
\end{equation}
where $g=det(g_{\mu\nu})$. Then the full action is the sum of the Einstein-Hilbert term and a matter (e.g. scalar fields) term, $\mathcal{L}_M=\frac{1}{2}\partial_\mu \phi\partial^\mu \phi-V(\phi)$,
\begin{equation}
\textsl{S}=\int\left[-\frac{1}{16\pi G} R+\mathcal{L}_M\right]\sqrt{-g}d^4x.
\end{equation}
Note that $\sqrt{-g}$ is introduced in the integration measure to make $\mathcal{L}\sqrt{-g}$ a scalar density, and gravity couples to matter minimally through this factor. Einstein gravity assumes this minimal coupling to ensure the equivalence principle.

According to the action principle, the variation of an action with respect to $g^{\mu\nu}$ is zero :
\begin{eqnarray}
0&=&\delta _{g^{\mu\nu}}\textsl{S}\\\nonumber
&=&\int\left[-\frac{1}{16\pi G}\frac{\delta(\sqrt{-g}R)}{\delta g^{\mu\nu}}+\frac{\delta(\sqrt{-g}\mathcal{L}_M)}{\delta g^{\mu\nu}}\right]\delta g^{\mu\nu}d^4x\\\nonumber
&=&\int\left[-\frac{1}{16\pi G}\left(\frac{\delta R}{\delta g^{\mu\nu}}+ \frac{R}{\sqrt{-g}}\frac{\delta(\sqrt{-g})}{\delta g^{\mu\nu}}\right)+\frac{1}{\sqrt{-g}} \frac{\delta(\sqrt{-g}\mathcal{L}_M)}{\delta g^{\mu\nu}}\right]\delta g^{\mu\nu}\sqrt{-g}d^4x.
\end{eqnarray}
As it is valid for any variation $\delta g^{\mu\nu}$, the square bracket has to vanish. Thus we obtain the following equation of motion
\begin{equation}
\label{2.22}
\frac{\delta R}{\delta g^{\mu\nu}}+ \frac{R}{\sqrt{-g}}\frac{\delta\sqrt{-g}}{\delta g^{\mu\nu}}=16\pi G\frac{1}{\sqrt{-g}} \frac{\delta(\sqrt{-g}\mathcal{L}_M)}{\delta g^{\mu\nu}}.
\end{equation}
We define the right hand side of the equation as an energy momentum tensor $T_{\mu\nu}$
\begin{equation}
\label{2.24}
T_{\mu\nu}\equiv\frac{2}{\sqrt{-g}} \frac{\delta(\sqrt{-g}\mathcal{L}_M)}{\delta g^{\mu\nu}}.
\end{equation}
From this equation of motion we derive the Einstein's field equation. To do that we need to know what $\delta R$ and $\delta\sqrt{-g}$ are. We start with the Riemann tensor in order to calculate $\delta R$.
\begin{equation}
\delta R^{\rho}_{\sigma\mu\nu}=\partial_\mu\delta\Gamma^{\rho}_{\nu\sigma}-\partial_\nu\delta\Gamma^{\rho}_{\mu\sigma}+\delta\Gamma^{\rho}_{\mu\lambda}\Gamma^{\lambda}_{\nu\sigma}+\Gamma^{\rho}_{\mu\lambda}\delta\Gamma^{\lambda}_{\nu\sigma}-\delta\Gamma^{\rho}_{\nu\lambda}\Gamma^{\lambda}_{\mu\sigma}-\Gamma^{\rho}_{\nu\lambda}\delta\Gamma^{\lambda}_{\mu\sigma}
\end{equation}
and the variation $\delta\Gamma^{\rho}_{\nu\mu}$ is the difference of two connections, therefore it is itself a tensor \cite{Carroll:2004st}. Thus its covariant derivative is given by 
\begin{equation}
\nabla_\lambda(\delta\Gamma^{\rho}_{\nu\mu})=\partial_\lambda(\delta\Gamma^{\rho}_{\nu\mu})+\Gamma^{\rho}_{\sigma\lambda}\delta\Gamma^{\sigma}_{\nu\mu}-\Gamma^{\sigma}_{\nu\lambda}\delta\Gamma^{\rho}_{\sigma\mu}-\Gamma^{\sigma}_{\mu\lambda}\delta\Gamma^{\rho}_{\nu\sigma}.
\end{equation}
From this expression we can directly read off
\begin{eqnarray}
\delta R^{\rho}_{\sigma\mu\nu}&=&\nabla_\mu(\delta\Gamma^{\rho}_{\nu\sigma})-\nabla_\nu(\delta\Gamma^{\rho}_{\mu\sigma}),\\\nonumber
\Rightarrow \delta R_{\mu\nu}&=&\nabla_\rho(\delta\Gamma^{\rho}_{\nu\mu})-\nabla_\nu(\delta\Gamma^{\rho}_{\rho\mu}).
\end{eqnarray}

The Ricci scalar is defined by  $R=g^{\mu\nu}R_{\mu\nu}$, then
\begin{eqnarray}
\delta R&=& R_{\mu\nu}\delta g^{\mu\nu}+g^{\mu\nu}\delta R_{\mu\nu}\\\nonumber
&=&R_{\mu\nu}\delta g^{\mu\nu}+\nabla_\rho g^{\mu\nu}\delta\Gamma^{\rho}_{\mu\nu}-\nabla_\nu g^{\mu\nu}\delta\Gamma^{\rho}_{\rho\mu}\\\nonumber
&=&R_{\mu\nu}\delta g^{\mu\nu}+\nabla_\sigma ( g^{\mu\nu}\delta\Gamma^{\sigma}_{\mu\nu}-g^{\mu\sigma}\delta\Gamma^{\rho}_{\rho\mu}).
\end{eqnarray}
The second term of the right hand side is a total derivative, and thus, according to Stoke's theorem, it can be expressed as boundary term by integral. As we assume the variation of the metric $\delta g^{\mu\nu}$ vanishes at infinity, this term does not contribute to the variation of the action. Therefore, we get the following expression
\begin{equation}
\frac{\delta R}{\delta g^{\mu\nu}}=R_{\mu\nu}.
\end{equation}

Now we consider $\delta\sqrt{-g}$. Since $\delta g=\delta det(g_{\mu\nu})$, according to Jacobi's formula\footnote{The Jacobi's formula is a rule for differentiating a determinant. If a matrix $A$ is invertible for all $t$, then $\frac{d}{dt}det  A(t)= det  A(t) tr \left(A(t)^{-1}\frac{d}{dt}A(t)\right)$.}, it can be rewritten as $g g^{\mu\nu}\delta g_{\mu\nu}$. Then it follows
\begin{equation}
\delta\sqrt{-g}=-\frac{1}{2\sqrt{-g}}\delta g=\frac{1}{2}\sqrt{-g}(g^{\mu\nu}\delta g_{\mu\nu}).
\end{equation}
Since $\delta(g^{\mu\nu}g_{\mu\nu})=0$, it follows that $g_{\mu\nu}\delta g^{\mu\nu}=-g^{\mu\nu}\delta g_{\mu\nu}$ and
$$
\delta\sqrt{-g}=-\frac{1}{2}\sqrt{-g}(g_{\mu\nu}\delta g^{\mu\nu}),
$$

\begin{equation}
\label{2.30}
\Rightarrow \frac{1}{\sqrt{-g}}\frac{\delta\sqrt{-g}}{\delta g^{\mu\nu}}=-\frac{1}{2}g_{\mu\nu}.
\end{equation}
The left hand side of the equation of motion (\ref{2.22}) finally becomes $R_{\mu\nu}-\frac{1}{2}g_{\mu\nu}R$ so that the equation of motion takes the form of Einstein's field equation\\
\begin{equation}
R_{\mu\nu}-\frac{1}{2}g_{\mu\nu}R=8\pi G T_{\mu\nu}.
\end{equation}

We come back to the energy momentum tensor
\begin{equation}
T_{\mu\nu}\equiv\frac{2}{\sqrt{-g}} \frac{\delta(\sqrt{-g}\mathcal{L}_M)}{\delta g^{\mu\nu}}.
\end{equation}
The variation with respect to $\delta g^{\mu\nu}$ splits into two parts, $\frac{2}{\sqrt{-g}} \frac{\delta\sqrt{-g}}{\delta g^{\mu\nu}}\mathcal{L}_M$ and $\frac{2}{\sqrt{-g}} \frac{\delta\mathcal{L}_M}{\delta g^{\mu\nu}}\sqrt{-g}$.\\
From the variation of the determinant (\ref{2.30}), we get

\begin{equation}
\frac{2}{\sqrt{-g}} \frac{\delta\sqrt{-g}}{\delta g^{\mu\nu}}\mathcal{L}_M=-g_{\mu\nu}\mathcal{L}_M.
\end{equation}
Now we assume a scalar field as the matter sector, i.e., $\mathcal{L}_M=\frac{1}{2}\partial_\mu \phi\partial^\mu \phi-V(\phi)$, then

\begin{equation}
2\frac{\delta\mathcal{L}_M}{\delta g^{\mu\nu}}=2\frac{\delta}{\delta g^{\mu\nu}}\left(\frac{1}{2}g^{\alpha\beta}\partial_\alpha\phi\partial_\beta\phi\right)-2\frac{\delta}{\delta g^{\mu\nu}}V(\phi).
\end{equation}
Hence $\frac{\delta g^{\alpha\beta}}{\delta g^{\mu\nu}}={\delta^\alpha}_\mu {\delta^\beta}_\nu$ and $V(\phi)$ being independent of the metric $g^{\mu\nu}$ leads to
\begin{eqnarray}
2\frac{\delta\mathcal{L}_M}{\delta g^{\mu\nu}}&=&\delta^\alpha_\mu \delta^\beta_\nu\partial_\alpha\phi\partial_\beta\phi\\\nonumber
&=&\partial_\mu\phi\partial_\nu\phi.
\end{eqnarray}
Now we have the energy momentum tensor in term of  the scalar field
\begin{equation}
T_{\mu\nu}=\partial_\mu\phi \partial_\nu\phi-g_{\mu\nu}\left[\frac{1}{2}\partial_\alpha\phi\partial^\alpha\phi-V(\phi)\right],
\end{equation}
or equally
\begin{equation}
{T^\mu}_\nu=\partial^\mu\phi \partial_\nu\phi-\left[\frac{1}{2}\partial_\alpha\phi\partial^\alpha\phi-V(\phi)\right]{\delta^\mu}_\nu.
\end{equation}

For the (0,0) component we get ${T^0}_0=\rho=\frac{1}{2}\dot{\phi}^2+V(\phi)$. For the (i,i) component we get $-\frac{1}{3}{T^{i}}_{i}=p=\frac{1}{2}\dot{\phi}^2-V(\phi)$.
If we take the FLRW metric and assume the Universe is homogeneous, we obtain following Friedmann equations
\begin{equation}
H^2+\frac{k}{a^{2}}=\frac{8\pi G}{3}\left(\frac{1}{2}\dot{\phi}^2+V(\phi)\right),
\end{equation}
\begin{equation}
\label{ }
\frac{\ddot{a}}{a}=-\frac{8\pi G}{3}\left(\dot{\phi}^2-V(\phi)\right).
\end{equation}

Let us check again the Lagrange density $\mathcal{L}$. The full Lagrange density including the gravity term and the scalar field term is
\begin{equation}
\mathcal{L}\sqrt{-g}=-\frac{1}{16\pi G}R\sqrt{-g}+\frac{1}{2}\partial_\mu\phi \partial^\mu\phi\sqrt{-g}-V(\phi)\sqrt{-g}.
\end{equation}
The functional derivative of $\mathcal{L}$ with respect to $\phi$ is zero according to Euler lagrange equation

\begin{eqnarray}
\label{2.46}
\frac{\delta\textsl{S}\left[g,\phi\right]}{\delta \phi}&=&\frac{\partial\mathcal{L}\sqrt{-g}}{\partial\phi}-\partial_\nu\frac{\partial\mathcal{L}\sqrt{-g}}{\partial(\partial^\nu\phi)}=-V'(\phi)\sqrt{-g}-\partial_\nu(\partial^\nu\phi\sqrt{-g})\\\nonumber
&=&-V'(\phi)\sqrt{-g}-(\partial_\nu\partial^\nu\phi)\sqrt{-g}-\partial^\nu\phi(\partial_\nu\sqrt{-g})=0,
\end{eqnarray}
where the prime denotes $\frac{d}{d\phi}$.

We take again the FLRW metric, and assume the homogeneous Universe to drop spatial gradients. For a diagonal matrix, the determinant is the product of the diagonal components
\begin{equation}
\label{2.49}
g=-\frac{a^6 r^4 sin^2\theta}{1-k r^2}.
\end{equation}
Because of spatial homogeneity $\partial^i\phi$ vanishes, and we obtain
\begin{equation}
\label{2.50}
0=-V'(\phi)\sqrt{-g}-\ddot{\phi}\sqrt{-g}-\dot{\phi}(\partial_t\sqrt{-g}).
\end{equation}
Since $\partial_t\sqrt{-g}=\frac{1}{2}\frac{-\dot{g}}{\sqrt{-g}}=\frac{3\dot{a}}{a}\sqrt{-g}$, we obtain
\begin{equation}
0=-V'(\phi)-\ddot{\phi}-\dot{\phi}\frac{3\dot{a}}{a},
\end{equation}
which yields the scalar field equation
\begin{equation}
\ddot{\phi}+3H\dot{\phi}+V'(\phi)=0.
\end{equation}
In the same way, we can derive the equations of motion from an action for each model.

\providecommand{\href}[2]{#2}\begingroup\raggedright\endgroup

\end{document}